\global\def\draftcontrol{0}

   \def\versionno{ n2* desitter}
\catcode`\@=11

\expandafter\ifx\csname draftcontrol\endcsname\relax\global\def\draftcontrol{0}
\fi

{\count255=\time\divide\count255 by 60
\xdef\hourmin{\number\count255}
\multiply\count255 by-60\advance\count255 by\time
\xdef\hourmin{\hourmin:\ifnum\count255<10 0\fi\the\count255}}
\def\draftdate{\number\month/\number\day/\number\year\ \ \ \hourmin }

\newcommand\makepapertitle{\par
  \begingroup
    \renewcommand\thefootnote{\@fnsymbol\c@footnote}%
    \def\@makefnmark{\rlap{\@textsuperscript{\normalfont\@thefnmark}}}%
    \long\def\@makefntext##1{\parindent 1em\noindent
            \hb@xt@1.8em{%
                \hss\@textsuperscript{\normalfont\@thefnmark}}##1}%
     \newpage
     \global\@topnum\z@   % Prevents figures from going at top of page.
     \@makepapertitle
     \thispagestyle{empty}\@thanks
  \endgroup
  \setcounter{footnote}{0}%
  \global\let\thanks\relax
  \global\let\makepapertitle\relax
  \global\let\@makepapertitle\relax
  \global\let\@thanks\@empty
  \global\let\@author\@empty
  \global\let\@date\@empty
  \global\let\@title\@empty
  \global\let\title\relax
  \global\let\author\relax
  \global\let\date\relax
  \global\let\and\relax
  \def\version{\let\version\@version\@gobble}
}
\def\@makepapertitle{%
  \newpage
   \ifnum\draftcontrol=1 {}
   \version\versionno
   \vskip 3em%
   \else
   \hfill\hbox to 3cm {\parbox{4cm}{\@pubnum}\hss}%
   \vskip 3em%
   \fi
   \begin{center}%
   \let \footnote \thanks
     {\LARGE {\@title}}%
     \vskip 1.5em%
     {\normalsize%\large
       \lineskip .5em%
       \begin{tabular}[t]{c}%
         \@author
       \end{tabular}\par}%
     \vskip 1.5em%
     {\@bstract}%
     \end{center}%
     \vskip 1.5em
     \@date%
   \par
}

\gdef\@pubnum{}
\def\pubnum#1{%
  \gdef\@pubnum{#1}}

\gdef\@bstract{}
\def\Abstract#1{%
  \gdef\@bstract{%
   \parbox{\textwidth-0pc}{%
   \centerline{\bf Abstract}\penalty1000%
\kern.2cm%
\noindent%\abstractfont \baselineskip=12pt
\renewcommand\baselinestretch{1.0}%
{#1}}}
}

\def\ps@paper{\let\@mkboth\@gobbletwo%
     \ifnum\draftcontrol=1
    \def\@oddfoot{\hbox to \textwidth{\tiny \versionno \hfil\tiny\draftdate}%
    \hskip -\textwidth \hbox to \textwidth{\hfil\rm\thepage\hfil}}%
     \else\def\@oddfoot{\hbox to \textwidth{\hfil\rm\thepage\hfil}}
     \fi
     \let\@evenfoot\@oddfoot
}
\def\body{\clearpage
          \pagestyle{paper}
    }
\def\@version#1{\ifnum\draftcontrol=1
\typeout{}\typeout{#1}\typeout{}
\vskip3mm\centerline{\hbox{\fbox{\normalsize{\tt DRAFT -- #1 -- }
                   {\draftdate}}}}\vskip3mm
\fi}
\let\version\@version
\long\def\eqlabel#1{\ifnum\draftcontrol=1
                    \tag@false  % there are some problems with multline without this
                    \tag*{(\theequation) \hbox to -0.2cm{\hspace{0cm}\small{#1}\hss}}
                    \refstepcounter{equation}
                    \edef\@currentlabel{\theequation}
                    \ltx@label{#1}          % use old LaTeX \label instead of new definition
                    \else
                    \label{#1}
                    \fi
                    }
\let\st@bibitem\@bibitem
\let\st@lbibitem\@lbibitem
\ifnum\draftcontrol=1
  \def\@bibitem#1{%
    \st@bibitem{#1}\a@@label{#1}\ignorespaces}
  \def\@lbibitem[#1]#2{%
    \st@lbibitem[#1]{#2}\a@@label{#2}\ignorespaces}
  \def\a@@label#1{%
    \gdef\a@lab{\smash{\normalfont\small#1}}
    \ifvmode
      \if@inlabel
        \global\setbox\@labels\hbox{%
          \llap{\a@lab\let\a@lab\relax
                \kern\@totalleftmargin\kern\marginparsep}%
          \box\@labels}%
      \fi
    \fi}
\fi
\documentclass[12pt,letterpaper]{article}
\usepackage{amsmath,amssymb,array,calc,epsfig,rotating,bm,xcolor}
\usepackage[sort]{cite}
\usepackage{graphicx}
\usepackage{psfrag,verbatim}

\ifnum\draftcontrol=1
\fi
\renewcommand\baselinestretch{1.25}
\renewcommand\section{\@startsection {section}{1}{\z@}%
                                   {-3.5ex \@plus -1ex \@minus -.2ex}%
                                   {2.3ex \@plus.2ex}%
                                   {\normalfont\large\bfseries}}
\renewcommand\subsection{\@startsection{subsection}{2}{\z@}%
                                   {-3.25ex\@plus -1ex \@minus -.2ex}%
                                   {1.5ex \@plus .2ex}%
                                   {\normalfont\normalsize\bfseries}}
\renewcommand\subsubsection{\@startsection{subsubsection}{3}{\z@}%
                                   {-3.25ex\@plus -1ex \@minus -.2ex}%
                                   {1.5ex \@plus .2ex}%
                                   {\normalfont\normalsize\it}}
\renewcommand\paragraph{\@startsection{paragraph}{4}{\z@}%
                                   {-3.25ex\@plus -1ex \@minus -.2ex}%
                                   {1.5ex \@plus .2ex}%
                                   {\normalfont\normalsize\bf}}

\numberwithin{equation}{section}

\def\revise#1       {\raisebox{-0em}{\rule{3pt}{1em}}%
                     \marginpar{\raisebox{.5em}{\vrule width3pt\
                     \vrule width0pt height 0pt depth0.5em
                     \hbox to 0cm{\hspace{0cm}{%
                     \parbox[t]{4em}{\raggedright\footnotesize{#1}}}\hss}}}}

\newcommand\nxt[1]  {\\\fnxt#1}
\newcommand{\ie}{{\it i.e.,}\ }
\newcommand{\eg}{{\it e.g.,}\ }

\def\cale         {{\cal E}}
\def\calf         {{\cal F}}

\def\cali         {{\cal I}}

\def\call         {{\cal L}}
\def\calm         {{\cal M}}
\def\caln         {{\cal N}}
\def\calo         {{\cal O}}

\def\calr         {{\cal R}}
\def\cals         {{\cal S}}

\def\zet          {{\mathbb Z}}

\def\del          {\partial}

\def\tr           {\mathop{\rm Tr}}

\def\Im           {{\rm Im\hskip0.1em}}

\def\sqr#1#2{{\vcenter{\vbox{\hrule height.#2pt
 \hbox{\vrule width.#2pt height#1pt \kern#1pt
 \vrule width.#2pt}\hrule height.#2pt}}}}

\def\a{\alpha}

\def\r{\rho}
\def\dd{\delta}

\def\aa1{\phi}
\def\cc1{\psi}

\def\k{\kappa}

\def\l{\lambda}

\def\k{\kappa}

\catcode`\@=12

\begin{document}

%%%
%%%%%% text starts here
%%%%%%%%%

\title{\bf Entanglement entropy of ${\cal N}=2^*$ de Sitter vacuum}

\date{September 10, 2019}
%\date\today

\author{
Alex Buchel\\[0.4cm]
\it $ $Department of Applied Mathematics\\
\it $ $Department of Physics and Astronomy\\ 
\it University of Western Ontario\\
\it London, Ontario N6A 5B7, Canada\\
\it $ $Perimeter Institute for Theoretical Physics\\
\it Waterloo, Ontario N2J 2W9, Canada
}

\Abstract{de Sitter vacuum of nonconformal gauge theories is non-equilibrium,
manifested by a nonvanishing rate of the comoving entropy production
at asymptotically late times. This entropy production rate is related
to the entanglement entropy of the de Sitter vacuum of the theory. We
use holographic correspondence to compute vacuum entanglement entropy
density $s_{ent}$ of mass deformed ${\cal N}=4$ supersymmetric
Yang-Mills theory --- the ${\cal N}=2^*$ gauge theory --- for various
values of the masses and the coupling constant to the background
space-time curvature. For a particular choice of the curvature
coupling, the Euclidean model can be solved exactly using the
supersymmetric localization.  We show that ${\cal N}=2^*$ de Sitter
entanglement entropy is not the thermodynamic entropy of the
localization free energy at de Sitter temperature.  Neither it is
related to the thermal entropy of de Sitter vacuum of pair-produced
particles.}

\makepapertitle

\body

\version\versionno
\tableofcontents

\section{Introduction}\label{intro}
The fundamental problem is understanding the current accelerated expansion of the
Universe \cite{Riess:1998cb,Perlmutter:1998np}. A positive cosmological constant
is the most straightforward explanation, however it implies that we live in (asymptotically)
de Sitter space-time. Whether or not de Sitter vacua are consistent in quantum gravity (String Theory)  is
an active area of research \cite{Kachru:2003aw,Obied:2018sgi}.

In this paper we do not study quantum de Sitter gravity, rather, we
focus on a much simpler problem --- unambiguous\footnote{A vacuum
energy of a QFT in $dS_4$ is {\it not} unambiguous --- it is
renormalization scheme dependent, which is one facet of the
cosmological constant problem.} signatures of a QFT in classical de
Sitter background space-time.  An observable of interest here is the
entropy density of a QFT in de Sitter vacuum. Consider a typical
state of a theory specified by a density matrix $\rho$. Given
the density matrix, we associate the von Neumann entropy ${\cals}$,
\begin{equation}
\cals=-\tr(\r \ln \r)\,.
\eqlabel{svn}
\end{equation}
An initial state $\r(t=0)$ would evolve with time, driven in particular by  the
accelerated expansion of the background space-time of the theory.
As a result, the von Neumann entropy will be time-dependent.
According to the second law of thermodynamics,
the entropy is non-decreasing during the evolution
\begin{equation}
\frac{d}{dt}\ \cals \ \ge\ 0\,.
\eqlabel{svn1}
\end{equation}
If the evolution is adiabatic at late times, \ie the system reaches the equilibrium,
the entropy production rate vanishes.
In what follows, we are interested in the rate of the entropy production
at asymptotically late times in interacting gauge theories in de Sitter
space time.

It is difficult, if impossible, to address the posed question directly
in the QFT framework.  For example, how precisely one would
coarse-grain the microstate of the interacting system to produce
initial density matrix $\rho$? Also, it is almost out of question to
perform reliable dynamics of the gauge theories at finite gauge
coupling. Remarkably, these obstacles are eliminated in the
holographic framework \cite{Maldacena:1997re,Aharony:1999ti}.  In the
context of Maldacena duality, a strongly coupled gauge theory at
(infinitely) large 't Hooft coupling and in the planar limit is dual
to a gravitational bulk geometry. Although the precise dictionary is
unknown, a well-defined initial state on the gravitational side is dual
to  some coarse-grained state\footnote{Exactly how the initial state
on the gravitational side encodes the coarse-graining in the QFT
language is unknown.}  on the gauge theory side. We emphasize that
a generic initial state is necessarily mixed, because the gravitational
dual has an apparent horizon
\cite{Chesler:2013lia,Booth:2005qc,Figueras:2009iu}
(unless the initial state is fine-tuned --- typically a supersymmetric
equilibrium state).  It is natural to associate the coarse-grained
entropy $\cals$ of the boundary gauge theory with the gravitational
entropy of the apparent horizon \cite{Booth:2005qc,Figueras:2009iu}.
For spatially extended horizons, as under the discussion below, it is more
convenient to talk about the entropy density $s$ rather than the full
entropy $\cals$. Gauge theory dynamics is encoded in the gravitational
bulk evolution. It is straightforward to answer the question regarding
the late-time entropy production rate of the gauge theory in de Sitter
by simply following the dynamics of the apparent horizon of its
gravitational dual.

We begin reviewing the
argument presented in \cite{Buchel:2017pto} that a non-conformal gauge theory in de
Sitter space-time (flat or closed spatial slicing),
\begin{equation}
ds_4^2=-dt^2 +e^{2Ht}\ d\boldsymbol{x}^2\qquad  {\rm or}\qquad ds_4^2=-dt^2 +\frac{1}{H^2}\cosh^2(Ht)\ (dS^3)^2\,,
\eqlabel{slice}
\end{equation}
approaches at late-times a non-equilibrium vacuum state, characterized by a constant rate $\calr$ of the comoving
entropy production
\begin{equation}
\lim_{t\to\infty}\ \frac{1}{H^3a^3}\frac{d}{dt}(a^3 s)\equiv 3H\times \calr\,,
\eqlabel{rate}
\end{equation}
where $a=e^{Ht}$ or $a=\cosh(Ht)/H$ is the  scale factor. The rate $\calr$, unlike the stress-energy tensor of the theory,
is renormalization
scheme unambiguous. It depends on QFT scales (physical mass parameters and relevant renormalizable coupling constants)
breaking the conformal invariance; furthermore, in the limit where the scale invariance is restored,
\begin{equation}
\lim_{QFT\to CFT} \calr = 0\,,
\eqlabel{cftlimit}
\end{equation}
in agreement with the adiabaticity of the  (Euclidean) de Sitter CFT vacuum. Eq.~\eqref{rate}
suggests a simple physical meaning of the rate $\calr$ \cite{Buchel:2017qwd}:
\begin{equation}
\lim_{t\to \infty} s\equiv s_{ent}=H^3\ \calr\,.
\eqlabel{s}
\end{equation}
Even though the late-time de Sitter state can be assigned (from the holographic perspective) a Hawking temperature
\begin{equation}
T_{dS}=\frac{H}{2\pi}\,,
\eqlabel{tempdS}
\end{equation}
$s_{ent}$ can not be a thermal entropy of the QFT at $T=T_{dS}$. This is evident from the fact (using \eqref{cftlimit})
that\footnote{We return to this in section \ref{discussion}.}
\begin{equation}
s_{thermal}^{CFT}\bigg|_{T=T_{dS}}\propto H^3\qquad {\rm while}\qquad s_{ent}^{CFT}\propto \calr^{CFT}=0\,.
\eqlabel{thermalent}
\end{equation}

In this paper we focus on a precise holographic correspondence between
$\caln=2^*$ gauge theory and Pilch-Warner (PW) geometry of type IIB
supergravity \cite{Pilch:2000ue,Buchel:2000cn,Evans:2000ct}. There are
two reasons for this particular choice:
\nxt $\caln=2^*$ theory is a deformation of $\caln=4$ $SU(N)$ supersymmetric Yang-Mills theory.
Specifically, in the $\caln=2$ language\footnote{We use notations
of \cite{Bobev:2013cja}.}, the field content of the latter includes a
vector multiplet (a gauge field $A_\mu$, two Weyl fermions $\psi_1$ and
$\psi_2$ and a complex scalar $Z_3$) and a hypermultiplet consisting
of two complex scalar fields $Z_1$ and $Z_2$ and two Weyl fermions
$\chi_1$ and $\chi_2$. The mass deformation that results in
$\caln=2^*$ gauge theory is a mass term for the hypermultiplet:
\begin{equation}
\call_{\caln=2^*}=\call_{\caln=4}+m^2\ \tr\left(|Z_1|^2+|Z_2|^2\right)
+m\ \tr \left(\chi_1\chi_1+\chi_2\chi_2+{\rm h.c.}\right)\,.
\eqlabel{massdef}
\end{equation}
As a result, as it was already shown in \cite{Buchel:2017pto}, 
\begin{equation}
\calr^{\caln=2^*}\ne 0\,.
\eqlabel{rn2}
\end{equation}
\nxt The second reason is an attempt to exploit the precision holography in a non-conformal
setting in the context of large-$N$ $\caln=2^*$ gauge theory \cite{Pestun:2007rz,Buchel:2013id,Bobev:2013cja,Chen:2014vka,Zarembo:2014ooa,Chen-Lin:2017pay,Bobev:2018hbq,Russo:2019lgq}. Notice that the Wick rotation  of the closed de Sitter geometry is that of the four-sphere
\begin{equation}
-dt^2+ \frac{1}{H^2}\cosh^2(Ht)\ (dS^3)^2\qquad \underbrace{\Longrightarrow}_{t\to \frac iH \theta}\qquad
\frac{1}{H^2}\ (dS^4)^2\,,
\eqlabel{de4sp}
\end{equation}
of radius $1/H$. While the Poincare supersymmetries are completely broken for $\call_{\caln=2^*}$
model on $S^4$, $\caln=2$ (Euclidean) supersymmetry can be restored with appropriately tuned
curvature coupling\footnote{The $S^4$ supersymmetry is unique up to a discrete choice represented
by $k\to -k$.} $k$ \cite{Pestun:2007rz}:
\begin{equation}
\call_{\caln=2^*}^{(m,k)}\equiv\call_{\caln=2^*}+k\ \tr\left(Z_1^2+Z_2^2+{\rm h.c.}\right)\,,\qquad
k=\frac i2 m H\,.
\eqlabel{deflambda}
\end{equation}
Moreover, the model $\call_{\caln=2^*}^{(m,i m H/2)}$ on $S^4$ upon supersymmetric localization
is mapped to a zero dimensional matrix model  \cite{Pestun:2007rz}, which can be solved analytically
in the large-N limit \cite{Buchel:2013id,Chen:2014vka,Zarembo:2014ooa,Chen-Lin:2017pay,Russo:2019lgq},
allowing for  precision tests of the holographic correspondence. As we already mentioned,
the holographic dual to $\call_{\caln=2^*}^{(m,0)}$ is a PW type IIB supergravity --- the latter can not
be a holographic dual to $\call_{\caln=2^*}^{(m,k\ne 0)}$: explicit computations of the holographic and
the matrix model partition functions on $S^4$ demonstrated a disagreement \cite{Buchel:2013fpa}.
Shortly after the disagreement was pointed out, the correct holographic dual to $\call_{\caln=2^*}^{(m,k)}$
was identified \cite{Bobev:2013cja} (BEFP)\footnote{See \cite{Bobev:2018hbq} for a 10d uplift
of BEFP geometry.}. Of course, on the level of the effective actions, we have a consistent truncation
\begin{equation}
{\rm BEFP}\bigg|_{k=0}\ =\ {\rm PW}\,.
\eqlabel{limpw}
\end{equation}
The previous computation of the rate \eqref{rn2} was performed at $k=0$, and thus there is no
chance of understanding this quantity from the matrix model perspective.
This paper is an extension of the computations of \cite{Buchel:2017pto}
to $\calr^{\caln=2^*}_{(m,k)}$.

We want to stress the following facts:
\begin{itemize}
\item We do not understand how to compute $\calr^{\caln=2^*}_{(m,i m H/2)}$ from localization.
Neither do we understand the physical origin of the quantity\footnote{See section
\ref{discussion} for some comments.}.
We present the holographic computations of the quantity  $\calr^{\caln=2^*}_{(m,k)}$ in three cases:
\begin{equation}
\calr^{\caln=2^*}_{(m=\mu,k=i\mu H/2)}\,,\qquad \calr^{\caln=2^*}_{(m=i\mu,k=\mu H/2)}\,,
\qquad  \calr^{\caln=2^*}_{(m=\mu,k=\mu H/2)}\,,
\eqlabel{casesr}
\end{equation}
for a real mass parameter $\mu$.
\item Notice that the first two examples are related by an analytical continuation of the mass parameter
$\mu$, and we hope might be accessible from the matrix model. It is important to emphasize --- as we review
in details in section \ref{rn4} --- that the entropy density  $s_{ent}$ \eqref{s} appears to be 
(from the holographic perspective) an intrinsically Lorentzian quantity. This is in contrast to a
thermal entropy which can be understood holographically both in Lorentzian and Euclidean signatures
(\eg as an area of the dual black brane event horizon in Eddington-Finkelstein or (analytically continued)
Schwarzschild coordinates). 
\end{itemize}

Given \cite{Buchel:2017pto}, the results presented here are not conceptually new and are technical in nature.
Thus, we try to allocate to appendices as much details as possible, leaving only the physical aspects.
In section \ref{rn4} we compute $\calr$ for $\caln=4$ SYM. Of course the result is ``zero'' for this theory,
but it illustrates the set of holographic tools used in computing $\calr$ for more general models.
In section \ref{main} we compute \eqref{casesr}. Since we will perform the computations in a
different framework\footnote{The computational framework developed here is indispensable in
understanding the "phase diagram'' of late-time de Sitter states in confining models with
spontaneous symmetry breaking, such as Klebanov-Strassler gauge theory \cite{Klebanov:2000hb}.}
from the one used in  \cite{Buchel:2017pto}, an agreement
\begin{equation}
\calr^{\caln=2^*}=\calr^{\caln=2^*}_{(m,0)}
\eqlabel{match}
\end{equation}
is an important test. Finally, in section \ref{discussion} we present some comments on failed
interpretations of $s_{ent}$.

\section{$\calr^{\caln=4}=\calr^{\caln=2^*}_{(0,0)}=0$}\label{rn4}

The purpose of this section is to set up/review holographic tools used to compute $\calr$. We will do it in the simplest
context possible, \ie $\caln=4$ SYM, which allows for a completely analytic discussion.
The price we pay is a trivial result,
\begin{equation}
\calr^{\caln=4}=0\,.
\eqlabel{n4result}
\end{equation}

Consider a five-dimensional gravitational dual to to $\caln=4$ $SU(N)$ SYM\footnote{To further simplify the
discussion we will use a consistent truncation to the metric sector only. The discussion is readily extended in
the presence of the bulk scalar fields, dual to $\caln=4$ gauge invariant operators, see below.
The conclusion is unchanged.}:
\begin{equation}
S_{\caln=4}=\frac{1}{16\pi G_5}\int_{\calm_5} d^5\xi\sqrt{-g}\left[R+\frac{12}{L^2}\right]\,.
\eqlabel{n4action}
\end{equation}
In what follows we set\footnote{This is done to agree with the conventions in section \ref{main}.}
the asymptotic $AdS_5$ radius $L=2$, leading to the identification
\begin{equation}
G_5=\frac{4\pi}{N^2}\,.
\eqlabel{defgt5}
\end{equation}
A generic state of the gauge theory, homogeneous and isotropic in the spatial
boundary coordinates $\boldsymbol{x}=\{x,y,z\}$, leads to a dual gravitational metric ansatz
\begin{equation}
ds_5^2=2 dt\ (dr -A dt) +\Sigma^2\ d\boldsymbol{x}^2\,,
\eqlabel{EFmetric}
\end{equation}
with the warp factors $A,\Sigma$ 
depending only on $\{t,r\}$.
Notice that the metric \eqref{EFmetric} is invariant under the residual diffeomorphisms
as $r\to \bar{r}=r-\lambda(t)$
\cite{Chesler:2013lia}
\begin{equation}
A(t,r)\to \bar{A}(t,\bar{r})= A(t,r+\lambda(r))-\dot{\lambda}(t)\,,\qquad
\Sigma(t,r)\to \bar{\Sigma}(t,\bar{r})=\Sigma(t,r+\lambda(t))\,.
\eqlabel{invmetric}
\end{equation}
From
the effective action \eqref{n4action} we obtain the following equations of
motion:
\begin{equation}
\begin{split}
&0=\left(d_+\Sigma\right)'+2 {\Sigma'}\ d_+\ln\Sigma-
\frac \Sigma2,\\ 
&0=A''-6(\ln\Sigma)'\ d_+\ln\Sigma +\frac 12,
\end{split}
\eqlabel{ev1}
\end{equation}
as well as the Hamiltonian and the momentum constraint equations:
\begin{equation}
0=\Sigma''\,,\qquad 0=d^2_+\Sigma -2 A\Sigma' -(4 A \Sigma'+A' \Sigma)d_+\ln\Sigma 
+ \Sigma A \,.
\eqlabel{hammom}
\end{equation}
In \eqref{invmetric}-\eqref{hammom} 
we denoted $'= \frac{\del}{\del r}$,  $\dot\ =\frac{\del}{\del t}$, 
and $d_+= \frac{\del}{\del t}+A \frac{\del }{\del r}$. 
Assuming the SYM background metric (for now we keep the scale factor $a(t)$ arbitrary)
\begin{equation}
ds_4^2=-dt^2+a(t)^2\ d\boldsymbol{x}^2\,,
\eqlabel{flrw}
\end{equation}
the bulk metric \eqref{EFmetric} has the near-boundary $r\to\infty$ asymptotic behavior
\begin{equation}
\Sigma=\frac{ar}{2}+\calo(r^{0})\,,\qquad A=\frac{r^2}{8}+\calo(r^1)\,.
\eqlabel{bcdata}
\end{equation}
The most general solution to \eqref{ev1}-\eqref{hammom}, subject to the boundary conditions
\eqref{bcdata} takes form
\begin{equation}
\begin{split}
&A=\frac{(r+\lambda)^2}{8}-(r+\lambda)\ \frac{{\dot a}}{a}-{\dot\lambda}-\frac{r_0^4}{8a^4(r+\lambda)^2}\,,
\qquad \Sigma=\frac{(r+\lambda)a}{2}\,,
\end{split}
\eqlabel{n4solve}
\end{equation}
where $\lambda(t)$ is an arbitrary function, as in \eqref{invmetric}, and $r_0$ is a constant.
Without loss of generality we now set $\lambda(t)\equiv 0$.

The bulk metric \eqref{EFmetric} has an apparent horizon (AH)\footnote{In general AH
is observer dependent. It is natural to definite AH with respect to an observer reflecting the
symmetries of the spatial slices --- homogeneity and isotropy in $\boldsymbol{x}$ here.
Note that as $t\to \infty$, which is the limit where we define
the entropy production rate $\calr$ \eqref{rate},
the AH coincides with the event horizon.} at $r=r_{AH}$ where \cite{Chesler:2013lia} 
\begin{equation}
d_+\Sigma\bigg|_{r=r_{AH}}=0\qquad \Longrightarrow\qquad   r_{AH}=\frac{r_0}{a(t)}\,.
\eqlabel{ah}
\end{equation}
Following \cite{Booth:2005qc,Figueras:2009iu} we associate the non-equilibrium  (comoving) entropy density $a^3 s$
of the SYM  with the Bekenstein-Hawking entropy density of the apparent horizon  
\begin{equation}
a^3 s =\frac{\Sigma^3}{4 G_5}\bigg|_{r=r_{AH}}= \frac{N^2r_0^3}{128\pi}\,.
\eqlabel{as}
\end{equation}
Further, from \cite{Fodor:1996rf} we identify the surface gravity $\kappa_{suf}$ of the dynamical AH:
\begin{equation}
\kappa_{suf}=A'\bigg|_{r=r_{AH}}=\frac{r_0}{2a(t)}-\frac {d}{dt}\ln a(t)\,.
\eqlabel{defkappa}
\end{equation}
For a stationary horizon the surface gravity is constant, and one identifies the Hawking temperature as
\begin{equation}
T=\frac{\kappa_{suf}}{2\pi}\,.
\eqlabel{hawt}
\end{equation}
Here, we are having a dynamical (non-equilibrium) horizon, and temperature is not well-defined --- instead we define
local temperature as
\begin{equation}\
T_{loc}(t)\equiv \lim_{\dot a\to 0} \frac{\kappa_{suf}}{2\pi}=\frac{r_0}{2a(t)}\equiv\frac{T_0}{a(t)}\,,
\eqlabel{tloc}
\end{equation}
where $T_0\equiv r_0/2$. Notice that if the metric \eqref{flrw} expansion rate is positive, the local temperature
redshifts as excepted for a thermal state of a conformal theory. While the comoving entropy density \eqref{as}
is time-independent, the physical entropy density "dilutes'':
\begin{equation}
s(t)=\frac{\pi^2}{2}N^2\ T_{loc}^3\,,
\eqlabel{entphysn4}
\end{equation}
again, as one would expect for a thermal state of $\caln=4$ SYM plasma in FLRW Universe \eqref{flrw}.
It is straightforward to compute the stress-energy momentum tensor; one finds for the energy density
$\cale$ and the pressure $P$ \cite{Buchel:2017pto}
\begin{equation}
\begin{split}
\cale(t)=\frac 38 \pi^2 N^2 T_{loc}^4+\frac{3N^2}{32\pi^2}\ \frac{(\dot a)^4}{a^4}\,,\qquad P(t)=\frac 13\cale(t)-\frac{N^2}{8\pi^2}\ \frac{(\dot a)^2\ddot a}{a^3}\,.
\end{split}
\eqlabel{epn4}
\end{equation}
As expected \cite{Buchel:2017pto}:
\nxt the stress-energy tensor is conserved as a consequence of the
bulk momentum constraint \eqref{hammom}
\begin{equation}
\frac{d\cale}{dt}+3\frac {\dot a}{a}\left(\cale+P\right)=0\,;
\eqlabel{epcons}
\end{equation}
\nxt the energy density and the pressure are just conformal transformations
of the equilibrium thermal state (with $T_{loc}\to T_0$), properly accounting for the conformal
anomaly;
\nxt the trace anomaly is
\begin{equation}
-\cale+3 P=\frac{N^2}{32\pi^2}\left(R_{\mu\nu}R^{\mu\nu}-\frac 13 R^2\right)=-\frac{3N^2}{8\pi^2}
\frac{(\dot a)^2\ddot a}{a^3}\,.
\eqlabel{tracean}
\end{equation}

We now discuss the late-time dynamics of the model in de Sitter space-time \eqref{slice},
from the SYM perspective, and from the bulk geometry perspective. Finally, we outline how the results
can be obtained bypassing the construction of dynamical solution and focusing on the late-time limit directly.
The latter makes an argument why \eqref{n4result} is true for more general states of
the SYM, \eg the states where some operators of the SYM (bulk scalars) are initially excited.

\subsection{$\caln=4$ SYM perspective}

We take
\begin{equation}
a(t)=e^{H t}\,.
\eqlabel{defa}
\end{equation}
The local temperature of the plasma is given by \eqref{tloc}, the entropy density is given by
\eqref{entphysn4}, the energy density and the pressure are given by \eqref{epn4}. At late times
we find
\begin{equation}
\begin{split}
\lim_{t\to \infty} \left\{
\begin{matrix}
{T_{loc}}\\
\cale\\
P\\
a^3 s
\end{matrix}
\right\}\qquad =\qquad
 \left\{
\begin{matrix}
0\\
\frac{3N^2 H^4}{32\pi^2}\\
-\frac{3N^2 H^4}{32\pi^2}\\
\frac{\pi^2}{2}N^2 T_0^3
\end{matrix}
\right\}\,.
\end{split}
\eqlabel{n4stuff}
\end{equation}
Thus, $\caln=4$ late-time vacuum state in de Sitter is characterized by a cosmological constant,
a vanishing local temperature, a constant comoving entropy density, and a vanishing physical entropy
density. As a result, 
\begin{equation}
\lim_{t\to\infty}\ \frac{1}{H^3a^3}\frac{d}{dt}(a^3 s)=\lim_{t\to\infty}\ \left(\frac{1}{H^3a^3}\ \times 0\right)=0
\qquad \Longrightarrow\qquad \calr^{\caln=4}=0\,.
\eqlabel{raten4res}
\end{equation}

\subsection{Holographic dual perspective}\label{split}

The bulk geometry is characterized by two warp factors $A$ and $\Sigma$ \eqref{EFmetric}.
Note that as $t\to \infty$ (in the $\l(t)\equiv 0$ gauge),
\begin{equation}
\begin{split}
\lim_{t\to \infty} A(t,r)\equiv A_v(r)=\frac r8 (r- 8 H)\,,\qquad \lim_{t\to \infty} \frac{\Sigma(t,r)}{a(t)}
\equiv \sigma_v(r)=\frac r2\,.
\end{split}
\eqlabel{warpsn4}
\end{equation}
One can either use the $t\to\infty$ limit of \eqref{ah}, or compute the location of the AH
directly in the "vacuum geometry''
\begin{equation}
ds_{5,vacuum}^2=2 dt\ (dr-A_v dt)+e^{2Ht} \sigma_v^2\ d\boldsymbol{x}^2\,,
\eqlabel{EFvacuum}
\end{equation}
to find 
\begin{equation}
r_{AH,vacuum}=0\,.
\eqlabel{ahvacuu}
\end{equation}
It is instructive to separate the range of the radial coordinate $r\in[r_{AH,vacuum},+\infty)$
into two subregions:  $r\in [0,8H]\bigcup [8 H,+\infty)$. Let's consider these subregions separately.
\begin{itemize}
\item $r\in[8 H,+\infty)$: \\
In this case $A_v\ge 0$. With the change of coordinates\footnote{This is simply a transformation
from the Eddington-Finkelstein to Fefferman-Graham coordinate system.}
\begin{equation}
t\to \tau+\frac{1}{2H}\ \ln\left(1-\frac{8H}{r}\right)\,,
\eqlabel{deftau}
\end{equation}
the vacuum gravitational dual geometry \eqref{EFvacuum} takes form
\begin{equation}
ds_{5,vacuum}^2=\frac{r(r-8H)}{4}\biggl(-d\tau^2+e^{2 H\tau} d\boldsymbol{x}^2 \biggr)+\frac{4dr^2}{r(r-8 H)}\,.
\eqlabel{FGvacuum}
\end{equation}
Further introducing
\begin{equation}
r=8H\cosh^2\frac\rho2\,,\qquad \rho\in[0,+\infty)\,,
\eqlabel{defrho}
\end{equation}
we arrive at
\begin{equation}
ds_{5,vacuum}^2=4\biggl[\ H^2\sinh^2\rho\ \biggl(-d\tau^2+e^{2 H\tau}\ d\boldsymbol{x}^2 \biggr)+d\rho^2\ \biggr]\,,
\eqlabel{FGvacuumrho}
\end{equation}
which is $dS_4$ slicing of the Poincare patch $AdS_5$ geometry  \cite{Buchel:2002wf}.
It is easy to see that one gets identical late-time bulk geometry for closed spatial slicing de Sitter
boundary \eqref{slice}, see  \cite{Buchel:2017pto}. In the latter case,
\begin{equation}
ds_{5,vacuum}^2=4\biggl[\ H^2\sinh^2\rho\ \biggl(-d\tau^2+\frac {1}{H^2}\cosh^2( H\tau)\ (dS^3)^2 \biggr)+d\rho^2\ \biggr]\,,
\eqlabel{vacuumclosed}
\end{equation}
which upon analytical continuation $\tau\to\tau_E=i\theta/H$ becomes
\begin{equation}
ds_{5,vacuum,E}^2=4\biggl[\ \sinh^2\rho\ (dS^4)^2 +d\rho^2\ \biggr]\,.
\eqlabel{vacuumclosed2}
\end{equation}
There can not be any entropy assigned to \eqref{FGvacuumrho}-\eqref{vacuumclosed2}.
From the periodicity $\theta \sim \theta + 2\pi$ we can formally assign de Sitter temperature \eqref{tempdS}.
\item $r\in[0,8 H]$:\\
Here, $A_v\le 0$; this part of the geometry is completely invisible from the
(Euclidean) de Sitter perspective, as in early studies of the holographic duality
of strongly coupled gauge theories in de Sitter background space-time
\cite{Buchel:2001iu,Buchel:2002wf,Buchel:2002kj,Buchel:2004qg,Buchel:2013dla}. It is ``discovered''
by asking a dynamical question, \ie what is the evolution of the holographic gauge theory
in de Sitter\footnote{Ref.~\cite{Buchel:2017lhu} provides an explicit model
of the dynamical evolution.}? The proper bulk coordinates to address this question are the Eddington-Finkelstein
ones, and in those coordinates the spatial location with 
\begin{equation}
A_v=0\qquad  \Longrightarrow\qquad r=8 H\qquad {\rm or}\qquad \rho=0  \,,
\eqlabel{ehorizon}
\end{equation}
is not special; in fact, because of the AH condition \eqref{ah}
\begin{equation}
{\rm AH}:\qquad 0=\del_t \Sigma + A \del_r \Sigma\qquad \underbrace{\simeq}_{t\to \infty}\qquad
a(t) \left(H\sigma_v+ A_v \sigma'_v\right)\,,
\eqlabel{locah2}
\end{equation}
the spatial location \eqref{ehorizon} is {\it outside} the AH, and so to reach the AH one needs to extend the
geometry for $0\le r< 8H$ until the condition \eqref{locah2} is satisfied.
Working directly in vacuum geometry of the $\caln=4$ SYM   \eqref{EFvacuum},  since
\begin{equation}
\sigma_v\bigg|_{r=r_{AH,vacuum}}=0\,,
\eqlabel{vanishs}
\end{equation}
the physical entropy density of the late-time de Sitter vacuum of the
dual $\caln=4$ SYM vanishes\footnote{This limit is better defined using the last equality in \eqref{as}. 
} (see \eqref{as}):
\begin{equation}
s_{ent}=\frac{1}{4G_5}\ \lim_{t\to \infty} \left(\frac{\Sigma^3}{a^3}\bigg|_{r=r_{AH}}\right)=  \frac{1}{4G_5}\ \sigma_v^3\bigg|_{r=r_{AH,vacuum}}=0\,.
\eqlabel{sphys}
\end{equation}
Notice that the surface gravity of the AH is negative
\begin{equation}
\kappa_{vacuum}=A_v'\bigg|_{r=r_{AH,vacuum}}=-H\,.
\eqlabel{kvacuum}
\end{equation}
\end{itemize}

\subsection{$\caln=4$ SYM de Sitter late-time vacuum}\label{secvac}

Previously, we obtained the bulk geometry \eqref{EFvacuum} dual to late-time de Sitter vacuum of $\caln=4$ SYM
plasma by constructing a dual to a dynamical evolution of the thermal state (see \eqref{EFmetric} and \eqref{n4solve})
and taking the $t\to \infty$ limit of the latter. The dynamical evolution can be technically quite involved, and
is in fact unnecessary if one is interested in the late-time vacuum. Following  \cite{Buchel:2017pto}
we introduce
\begin{equation}
\lim_{t\to \infty}\left\{A(r,t),\frac{\Sigma(t,r)}{a(t)}\right\}=\{A,\sigma\}_v(r)\,,
\eqlabel{vacdef2}
\end{equation}
and take the $t\to \infty$ limit of the equations \eqref{ev1}-\eqref{hammom} instead
\begin{equation}
\begin{split}
&0=\sigma_v'+\frac{\sigma_v}{2A_v}(H-A_v')\,,\qquad 0=A_v'' -6 A_v \left((\ln \sigma_v)'\right)^2-6 H(\ln\sigma_v)'+\frac 12\,,
\end{split}
\eqlabel{veomsn}
\end{equation}
\begin{equation}
\begin{split}
&0=\sigma_v''\,,\qquad
0=\frac{1}{4A_v}- \left((\ln\sigma_v)'\right)^2-\frac{\sigma_v'}{2\sigma_vA_v}(2H+A_v')+
\frac{H}{4A_v^2}(H-A_v')\,.
\end{split}
\eqlabel{veomsn2}
\end{equation}
Solution of  \eqref{veomsn}-\eqref{veomsn2} (up to a shift of the radial coordinate) is \eqref{warpsn4}.

The first advantage of this approach is the fact that to solve for the dual late-time vacuum geometry
one needs to solve the system of ODEs \eqref{veomsn}-\eqref{veomsn2},
instead of PDEs  (as in \eqref{ev1}-\eqref{hammom}), followed by $t\to\infty$ limit. The second advantage is
a conceptual one: the late time vacuum state is {\it universal} while the approach towards it depends on the
details of the dynamical evolution of the chosen initial state. Indeed, the thermal state evolution discussed
at the beginning of this section depends on a single parameter $r_0$,  which can be related to the  local
temperature $T_0$ of the plasma at $a(t=0)=1$, see \eqref{tloc}; alternatively, it defines the constant comoving
entropy density during the evolution (as in \eqref{as}). There is no memory of the $r_0$ in the energy, pressure
or the physical entropy density of the late-time vacuum SYM.  Notice that during the evolution of the thermal state,
stress energy tensor behaves as 
\begin{equation}
T_{\mu\nu}(t) -T_{\mu\nu}^{vacuum}\ \sim\ \frac{r_0^4}{a(t)^4}\to 0 \qquad {\rm as}\qquad t\to\infty \,.
\eqlabel{stressenergy}
\end{equation}
One expects\footnote{This is explicitly confirmed by the fact
that de Sitter conformal theory vacuum is recovered in the limit of vanishing of the relevant couplings
of the non-conformal QFT, see \cite{Buchel:2017pto,Buchel:2017lhu}
and the analysis of section \ref{main}.} that a more generic homogeneous and isotropic initial state of
the $\caln=4$ SYM with initially a non-vanishing expectation value of
an operator $\calo$ of a fixed scaling dimension $\Delta$
\begin{equation}
\calo^\Delta (t=0)=\calo_0^\Delta\ne 0\,,
\eqlabel{o1}
\end{equation}
would evolve as
\begin{equation}
\calo^\Delta (t)\ \sim\ \frac{\calo_0^\Delta}{a(t)^\Delta}\to 0 \qquad {\rm as}\qquad t\to\infty \,,
\eqlabel{o2}
\end{equation}
leading to the same late-time vacuum state as the one obtained from the evolution of the
thermal state. Notice that in this more general setting the comoving entropy density
would increase, saturating at a fixed constant leading to the same conclusion \eqref{n4result}.
This follows from the fact that such more general  de Sitter space-time dynamics is Weyl equivalent to
a thermalization dynamics in Minkowski space-time (see Appendix B of \cite{Buchel:2017pto}),
in particular, the total comoving entropy  produced in de Sitter dynamics is exactly the same as
the total entropy produced in corresponding thermalization process of the conformal SYM in
Minkowski space-time.

\section{$\calr^{\caln=2^*}_{(m,k)}$}\label{main}

In section \ref{rn4} we discussed the computation of $\calr^{\caln=4}$.
We begin here outlining the strategy of computing the vacuum comoving
entropy production rate $\calr$ for $\caln=2^*$ gauge theory for
select choices of the mass parameter $m$ (see \eqref{massdef})
and the background space-time curvature coupling constant
$k$ (see \eqref{deflambda}), \ie $\calr^{\caln=2^*}_{(m,k)}$.
We highlight the differences from the $\caln=4$ SYM case.
\begin{itemize}
\item We follow approach of section \ref{secvac} and after deriving
the general evolution equations for homogeneous and isotropic state
of the $\caln=2^*$ gauge theory in de Sitter
(the analogue of \eqref{ev1}-\eqref{hammom}) we take the late-time
limit to arrive at the vacuum equations (the analogue of \eqref{veomsn}
-\eqref{veomsn2}).
\item Motivated by the discussion in section \ref{split} we consider
two subregions of the gravitational bulk geometry: from the
asymptotic $AdS_5$ boundary to $A_v=0$, and from $A_v=0$ to the
AH specified by \eqref{locah2}. We use Fefferman-Graham
(Schwarzschild) coordinates in the former subregion --- this would allow
us to make contact with BEFP numerical solution \cite{Bobev:2013cja}.
For the latter subregion we use the Eddington-Finkelstein coordinates.
We implement transition between the two coordinate systems at $A_v=0$.
\item The comoving entropy density is computed as in \eqref{sphys}.
Albeit here, contrary to the $\caln=4$ SYM, we find that
\begin{equation}
\sigma_v\bigg|_{r=r_{AH,vacuum}}\ne 0\,,
\eqlabel{sne0}
\end{equation}
provided $m\ne 0$, resulting in nonvanishing  $\calr^{\caln=2^*}_{(m,k)}$. 
\item We mention in passing (see \cite{Buchel:2017pto,Buchel:2017lhu})
that the conformal limit $(m,k)\to (0,0)$ is
nonanalytic: while $\sigma_v$ evaluated at the AH horizon  vanishes
in this limit, it does so with a fractional power of the 
conformal symmetry breaking scale (\eg see eq.(3.1) in \cite{Buchel:2017lhu}).
\item As in \cite{Buchel:2017qwd}, the surface gravity of the
AH is universally given by \eqref{kvacuum} for arbitrary $(m,k)$, see \eqref{kuniv}.
\end{itemize}

\subsection{BEFP and PW effective actions}

Effective five-dimensional gravitational action, holographically dual to $\caln=2^*$ gauge theory 
is\footnote{See \cite{Bobev:2018hbq} for the 10d type IIB supergravity uplift.}
\begin{equation}
\begin{split}
S_{BEFP}=&\frac{1}{16\pi G_5}\int_{\calm_5}d^5\xi\sqrt{-g}\biggl[
R-\frac{12}{\eta^2}\ \del_\mu\eta\del^\mu\eta-\frac{4}{(1-z\bar{z})^2}
\ \del_\mu z\del^\mu \bar{z}-V_{BEFP}\biggr]\,,\\
V_{BEFP}=&-\frac{1}{\eta^4}-2\eta^2\frac{1+z\bar{z}}{1-z\bar{z}}-\frac{\eta^8}{4}\frac{(z-\bar{z})^2}{(1-z \bar{z})^2}\,,
\end{split}
\eqlabel{sbefp}
\end{equation}
where
\begin{equation}
z=z_1-i z_2\,,\qquad \bar{z}=z_1+i z_2\,,\qquad \eta=e^\a\,.
\eqlabel{defzzbeta}
\end{equation}
The bulk scalars $\{z_i,\a\}$ are dual to the operators $\{\calo_i,\calo_\a\}$, implementing the
\[
\call_{\caln=4}\ \to\  \call_{\caln=2^*}\ \to\ \call_{\caln=2^*}^{(m,k)}\]
deformations \eqref{massdef} and \eqref{deflambda}.
Explicitly,
\begin{equation}
\calo_\a\sim \tr\left(|Z_1|^2+|Z_2|^2\right)\,,\ \calo_2\sim \tr\left(\chi_1\chi_1+\chi_2\chi_2
+{\rm h.c.}\right)\,,\ \calo_1\sim \tr\left(Z_1^2+Z_2^2+{\rm h.c.}\right)\,.
\eqlabel{opcorr}
\end{equation}
A consistent truncation
\begin{equation}
z_1\equiv 0\,,
\eqlabel{btopw}
\end{equation}
followed by a field redefinition
\begin{equation}
z_2\equiv \tanh \chi\,,
\eqlabel{defchipw}
\end{equation}
results in PW effective action \cite{Pilch:2000ue}, implementing the deformation
\eqref{massdef}\footnote{The stability of PW$\subset$BEFP embedding for the
thermal states of $\caln=2^*$ plasma (with $k=0$) was studied in \cite{Balasubramanian:2013esa}.}.

We are interested in late-time vacuum states of strongly coupled $\caln=2^*$ gauge theory
in de Sitter. The gravity dual ansatz describing dynamical evolution of $SO(4)$-invariant
states is
\begin{equation}
ds_5^2=2 dt\left(dr -Adt\right)+\Sigma^2\ (dS^3)^2\,,
\eqlabel{efbmetric}
\end{equation}
where the metric warp factors $\{A,\Sigma\}$, and the bulk scalar $\{z_i,\eta\}$,
are functions of $(t,r)$. The equations of motion obtained from \eqref{sbefp}
are given in \eqref{evb}-\eqref{momb}. These equations has to be supplemented
with the asymptotic $r\to \infty$ boundary conditions implementing the
gauge theory background de Sitter space-time \eqref{slice}, and
the deformation coupling constants $(m,k)$:
\begin{equation}
\begin{split}
&\Sigma=\frac r2\ \frac 1H\cosh(Ht)+\calo(r^0)\,,\qquad
A=\frac {r^2}{8}+\calo(r^1)\,,\\
&\eta=1-\frac{8m^2\ln r}{3r^2}+\calo(r^{-2})\,,\qquad z_1=\frac{16k\ln r}{r^2}+\calo(r^{-2})\,,
\qquad z_2=\frac{2m}{r}+\calo(r^{-2})\,.
\end{split}
\eqlabel{bcbefp}
\end{equation}
To derive the late-time geometry dual to $\caln=2^*$ vacuum in de Sitter, we introduce following
\cite{Buchel:2017pto}
\begin{equation}
\lim_{t\to\infty}\left\{\eta(t,r)\,,\ z(t,r)\,,\ \bar{z}(t,r)\,,\ A(t,r)\,,\ \frac{H }{\cosh(Ht)}\Sigma(t,r)\right\}
=\left\{\eta,z,\bar{z},A,\sigma\right\}_v(r)\,.
\eqlabel{vacn2*}
\end{equation}
The full set of the vacuum equations is given in \eqref{veoms}-\eqref{veoms2}. The first of
the constraint equations in \eqref{veoms2}
\begin{equation}
0=\sigma_v'+\frac{\sigma_v}{2A_v}(H-A_v')\,,
\eqlabel{impconst}
\end{equation}
is very important. Given the location of the AH,
\begin{equation}
AH:\qquad (H\sigma_v+A_v\sigma'_v)\bigg|_{r=r_{AH,vacuum}}=0\,,
\eqlabel{ahloc}
\end{equation}
it implies that \cite{Buchel:2017qwd} 
\begin{equation}
A'_v\bigg|_{r=r_{AH,vacuum}}=-H\qquad \Longrightarrow\qquad \kappa_{vacuum}=-H\,,
\eqlabel{kuniv}
\end{equation}
\ie the surface gravity of the late-time AH is universal --- it does not depend on the
mass deformation parameters $(m,k)$. 

Generic values of $(m,k)$ completely break the (Euclidean) supersymmetry of the model ---
as a result the gravitational equations of motion \eqref{veoms}-\eqref{veoms2}
are second-order for scalars $\{\eta_v,z_v,\bar{z}_v\}$ (and of the first-order for the
metric warp factors $\{A_v,\sigma_v\}$).  
However, it is straightforward to verify the the following first-order
equations\footnote{These equations can be obtained from the
BPS eqs.~(3.20) and (3.26) of \cite{Bobev:2013cja} transforming to EF coordinate system
and setting $L=2$.}
\begin{equation}
\begin{split}
&0={z}_v'+\frac{3\eta_v'(1-z_v\bar{z}_v)(\eta_v^6({z}_v-\bar{z}_v)+2(z_v+\bar{z}_v))}{2\eta_v
(\eta_v^6(\bar{z}_v^2-1)+1+\bar{z}_v^2)}\,,\\
&0=\bar{z}_v'+\frac{3\eta_v'(1-z_v\bar{z}_v)(\eta_v^6(\bar{z}_v-{z}_v)+2(z_v+\bar{z}_v))}{2\eta_v
(\eta_v^6({z}_v^2-1)+1+{z}_v^2)}\,,\\
&0=\left(\eta_v'\right)^2-\frac{\eta_v^6(z_v^2-\bar{z}_v^2)^2}{144H^2(1-z_v\bar{z}_v)^4}\,,
\end{split}
\eqlabel{susyEF}
\end{equation}
an algebraic expression for $A_v$,
\begin{equation}
A_v=\frac{2H^2(1-z_v\bar{z}_{v})^2(\eta_v^6({z}_v^2-1)+1+{z}_v^2)
(\eta_v^6(\bar{z}_v^2-1)+1+\bar{z}_v^2)}{\eta_v^8(z_v^2-\bar{z}_v^2)^2}\,,
\eqlabel{susyav}
\end{equation}
and $\sigma_v$ determined from \eqref{impconst}, solve all the second-order equations \eqref{veoms}
and the second constraint in \eqref{veoms2}. The BPS-like equations \eqref{susyEF} constrain the
non-normalizable coefficients $(m,k)$ of the scalars $z_{i,v}$ as  
\begin{equation}
4 k^2+ H^2 m^2=0\,,
\eqlabel{susycond}
\end{equation}
which is just a (Euclidean) supersymmetry condition
\eqref{deflambda}.

In what follows we set
\begin{equation}
H=1\,.
\eqlabel{seth}
\end{equation}
The correct $H$-dependence can be recovered from the dimensional analysis. 

\subsection{$\calr^{\caln=2^*}_{(m=i\mu,k=\mu H/2)}$ and $\calr^{\caln=2^*}_{(m=\mu,k=i\mu H/2)}$}\label{mainmuimu}

Consider first
\begin{equation}
(m,k)\ =\ \left(i\mu\,,\ \mu/2\right)\,,\qquad \Im\mu=0\,.
\eqlabel{case1}
\end{equation}
Introduce a new radial coordinate $x\in (0,x^*]$ covering the first bulk subregion (see section \ref{split})
\begin{equation}
\frac{dr}{dx}=-\frac 2x ({2 A_v})^{1/2}\,,
\eqlabel{eftofg1}
\end{equation}
implementing  the transformation from EF \eqref{efbmetric} to FG coordinate system  
\begin{equation}
ds_5^2=2 A_v\biggl(-(d\tau)^2+\frac{1}{H^2}\cosh^2(H \tau) (dS^3)^2\biggr)+ \frac{4}{x^2}\ dx^2\,,\qquad
d\tau=dt+\frac{2^{1/2}}{x({A_v})^{1/2}}\ dx\,,
\eqlabel{deffg1}
\end{equation}
where $x^*$ is defined so that
\begin{equation}
A_v\bigg|_{x=x^*}=0\,.
\eqlabel{defxs}
\end{equation}
Resulting equations\footnote{As in \cite{Bobev:2013cja}, the bulk scalars
$z$ and $\bar{z}$ should be understood as being independent for ``supersymmetric'' RG flows.}
for $\{z_v,\bar{z}_v,\eta_v\}$ and $s_v$, defined as
\begin{equation}
\sigma_v=\frac 1x+s_v\,,
\eqlabel{defsv}
\end{equation}
are collected in \eqref{case1r1}.
They must be solved numerically subject to the following asymptotes:
\begin{itemize}
\item asymptotic $AdS$ boundary, \ie  $x\to 0_+$,
\begin{equation}
\begin{split}
&z_v=\mu x+x^2 (-2 \mu \ln x+v)+\calo(x^3\ln x)\,,
\\
&\bar{z}_v=-\mu x+x^2 (-2 \mu \ln x+v)+\calo(x^3\ln x)\,,\\
&\eta_v=1-\frac{\mu}{3} (2 \mu \ln x-\mu-v) x^2+\calo(x^4)\,,\qquad  s_v = 2+x \left(1+\frac{\mu^2}{3}\right)+\calo(x^2) \,,
\end{split}
\eqlabel{uv11}
\end{equation}
where $v=v(\mu)$ is a single UV parameter;
\item $A_v=0$, \ie  $y\equiv (x^*-x)\to 0_+$,
\begin{equation}
\begin{split}
&z_v=\sqrt{\frac{e_0^6-1}{e_0^6+1}}\left(1-\frac{e_0^2}{3(x^*)^2}y^2+\calo(y^3)\right)\,,\\
&\bar{z}_v=\sqrt{\frac{e_0^6-1}{e_0^6+1}} \left(\frac{e_0^6-2}{e_0^6+2}
-\frac{e_0^2 (11 e_0^{12}-20)}{15(x^*)^2 (e_0^6+2)^2} y^2+\calo(y^3)\right)\,,\\
&\eta_v=e_0-\frac{e_0^{12}-1}{27e_0^3 (x^*)^2} y^2+\calo(y^3)\,,\qquad
s_v=s_0-\frac{1}{(x^*)^2}y+\calo(y^2) \,,
\end{split}
\eqlabel{ir11}
\end{equation}
characterized 3 additional parameters $\{e_0,s_0,x^*\}$. 
Note that
\begin{equation}
2A_v=\frac{4}{(x^*)^2}\ y^2+\frac{4}{(x^*)^3}\ y^3+\frac{28 e_0^{12}+297 e_0^4+80}{81e_0^4 (x^*)^4}\ y^4+\calo(y^5)\,,
\eqlabel{avit}
\end{equation}
thus, both the bulk geometry and the scalars are smooth in the vicinity of $x^*$.  
\end{itemize}

\begin{figure}[t]
\begin{center}
\psfrag{x}{{$\mu$}}
\psfrag{y}{{$v(\mu)$}}
\psfrag{z}{{$\dd v(\mu)$}}
\includegraphics[width=2.6in]{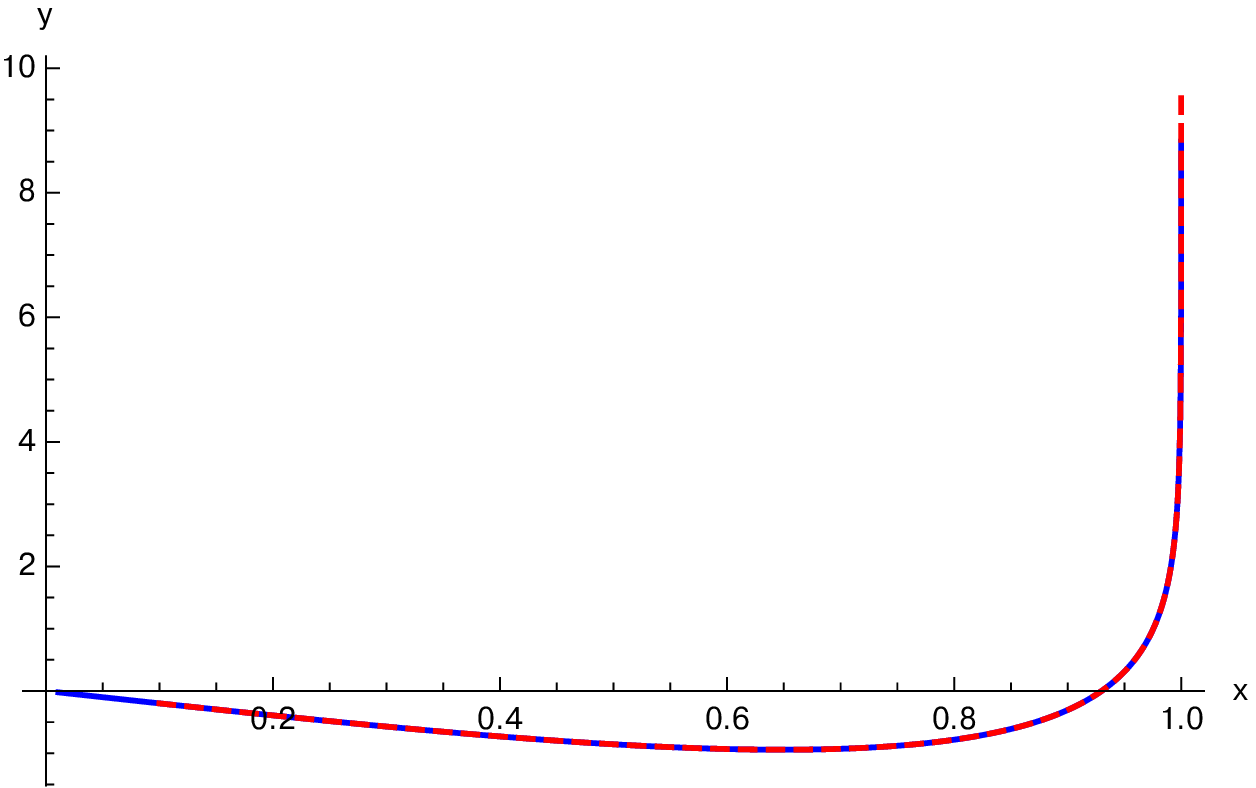}\qquad
\includegraphics[width=2.6in]{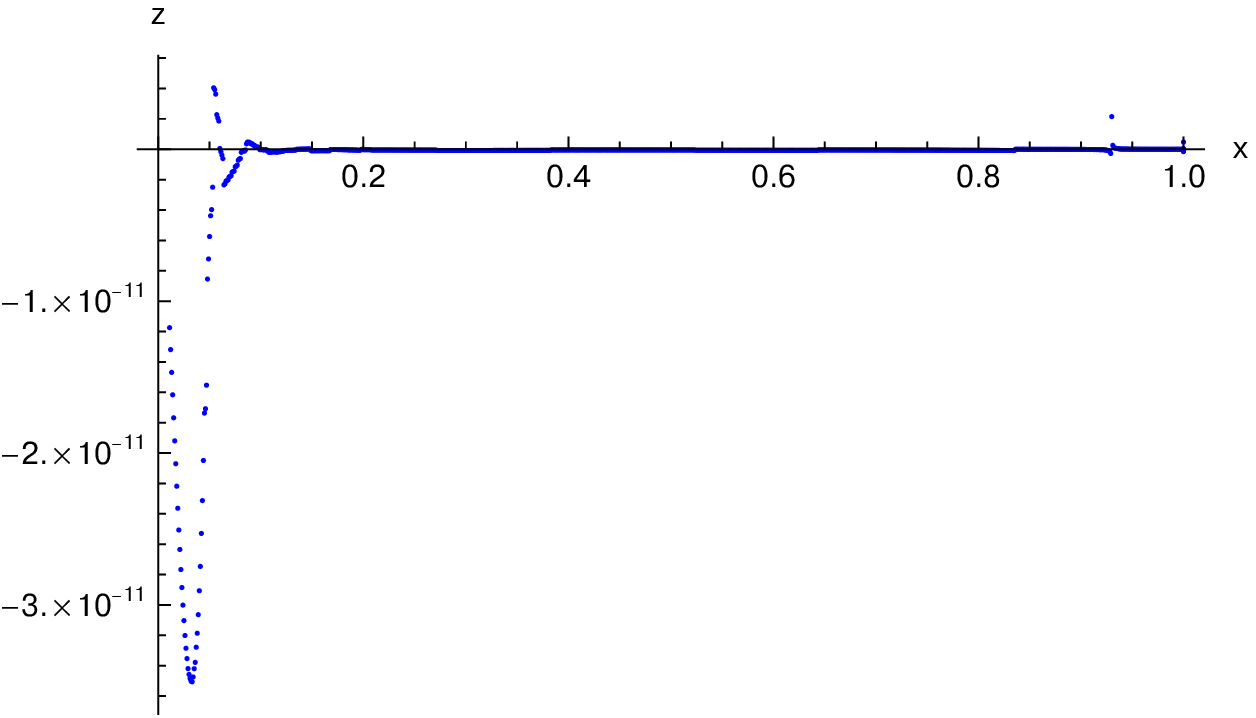}
\end{center}
  \caption{Left panel: numerical values for $v(\mu)$ (see \eqref{uv11}) (solid blue curve)
  and the predicted values $v_{prediction}$ (see \eqref{vpred}) (dashed red curve). Right panel:
  the residual $\dd v$ (see \eqref{defres}).
 } \label{figure1}
\end{figure}

Holographic renormalization of the model, along with the result for the
free energy computed from the localization \cite{Buchel:2013id}, makes a specific
prediction for $v(\mu)$  \cite{Bobev:2013cja}:
\begin{equation}
v\bigg|_{prediction}=-2\mu-\mu\ln(1-\mu^2)\,.
\eqlabel{vpred}
\end{equation}
Using the shooting method developed in \cite{Aharony:2007vg}, we recover numerical results
of  \cite{Bobev:2013cja}, and confirm the prediction \eqref{vpred}: fig.~\ref{figure1}
shows $v(\mu)$ (left panel, solid blue curve) and $v_{prediction}$
(left panel, dashed red curve) and the residual $\delta v$ (right panel)
\begin{equation}
\delta v\equiv 1-\frac{v(\mu)}{v_{prediction}}\,.
\eqlabel{defres}
\end{equation}

To compute the comoving entropy production rate $\calr^{\caln=2^*}_{(m=i\mu,k=\mu H/2)}$
we need an access to the second subregion of the bulk geometry with $A_v\le 0$,
see discussion in section \ref{split}. This is done returning to the original ER coordinate
$r$, see \eqref{eftofg1}. Introduce
\begin{equation}
r=r^*-\rho\,,\qquad \rho\in [0,\rho_{AH}]\ \Longleftrightarrow\ r\in[r_{AH,vacuum},r^*]\,,
\eqlabel{derrho}
\end{equation}
where 
\begin{equation}
r^*=r\bigg|_{x=x^*}\,.
\eqlabel{defrs}
\end{equation}
The holographic equations of motion in $\rho$ coordinate are simply \eqref{susyEF} with
\[
\del_r\ \equiv\ '\ =\ -\del_\rho\,.
\]
For small $\rho<0$ we obtain from the perturbative integration of \eqref{eftofg1}
\begin{equation}
-\rho=\frac{2}{(x^*)^2}\ y^2+\frac{2 }{(x^*)^3}\ y^3
+\frac{(7 e_0^{12}+297 e_0^4+20) }{162 e_0^4 (x^*)^4}\ y^4+\calo(y^5)\,,
\eqlabel{pertrho}
\end{equation}
or
\begin{equation}
\begin{split}
y=&\frac{(-\rho)^{1/2}x^*}{2^{1/2}} \biggl(
1-\frac{1}{2^{3/2}}\ (-\rho)^{1/2}-\frac{7 e_0^{12}-108 e_0^4+20}{1296e_0^4}\ (-\rho)
\\
&+\frac{2^{1/2}(7 e_0^{12}-27 e_0^4+20)}{2592e_0^4}\ (-\rho)^{3/2}+\calo\left((-\rho)^2\right)\biggr)\,,
\end{split}
\eqlabel{perty}
\end{equation}
which is used to set up the asymptotic initial conditions from \eqref{ir11}
\begin{equation}
\begin{split}
z_v=&\sqrt{\frac{e_0^6-1}{e_0^6+1}} \left(1+\frac16 e_0^2\ \rho+\frac{16 e_0^{12}+45 e_0^6+20}{3240e_0^2}\ \rho^2+\calo(\rho^3)\right)\,,
\\
\bar{z}_v=&\sqrt{\frac{e_0^6-1}{e_0^6+1)}} \biggl(\frac{e_0^6-2}{e_0^6+2}
+\frac{e_0^2 (11 e_0^{12}-20)}{30(e_0^6+2)^2}\ \rho\\
&+\frac{1304 e_0^{30}+10231 e_0^{24}+8750 e_0^{18}-10220 e_0^{12}-15400 e_0^6-5600}{113400e_0^2 (e_0^6+2)^3}\ \rho^2+\calo(\rho^3)\biggr)\,,
\\
\eta_v=&e_0+\frac{e_0^{12}-1}{54e_0^3}\ \rho+\frac{11 e_0^{24}-e_0^{12}-10}{6480e_0^7}\ \rho^2+\calo(\rho^3)\,,
\\
\sigma_v=&\frac{s_0 x^*+1}{x^*}-\frac{7 e_0^{12} s_0 x^*+7 e_0^{12}+20 s_0 x^*+20}{216x^* e_0^4}\
\rho\\
&- \frac{53 e_0^{24} s_0 x^*+53 e_0^{24}-133 e_0^{12} s_0 x^*-133 e_0^{12}+80 s_0 x^*+80}{29160e_0^8 x^*}\ \rho^2+\calo(\rho^3)\,.
\end{split}
\eqlabel{incondrho}
\end{equation}
Note that
\begin{equation}
A_v=-\rho+\frac{7 e_0^{12}+20}{216e_0^4}\ \rho^2+ \frac{53 e_0^{24}-133 e_0^{12}+80}{21870e_0^8}\ \rho^3+\calo(\rho^4)\,.
\eqlabel{avsin}
\end{equation}
Combining \eqref{impconst} and \eqref{ahloc} we have
\begin{equation}
\begin{split}
\frac{\sigma_v}{2}\left(H+A_v'\right)=
\frac{H(\eta_v^6(z_v^2-1)-2(z_v^2+1))(\eta_v^6(\bar{z}^2-1)+\bar{z}_v^2+1)\sigma_v}
{6\eta_v^2(z_v^2-\bar{z}_v^2)}
\bigg|_{r=r_{AH,vacuum}}=0\,,
\end{split}
\eqlabel{susyah}
\end{equation}
where in the second equality we used explicit expression for $A_v$ \eqref{susyav} and the
RG flow equations \eqref{susyEF}.
Eq.~\eqref{susyah} motivates introduction of the AH-monitoring function
\begin{equation}
Z_{AH}\equiv \frac{(\eta_v^6(z_v^2-1)-2(z_v^2+1))(\eta_v^6(\bar{z}^2-1)+\bar{z}_v^2+1)}
{6\eta_v^2(z_v^2-\bar{z}_v^2)}\,,\qquad Z_{AH}\bigg|_{\rho=\rho_{AH}}=0\,.
\eqlabel{zah}
\end{equation}
Interestingly, from \eqref{incondrho}
\begin{equation}
Z_{AH}=1-\frac{7 e_0^{12}+20}{216e_0^4} \rho-\frac{53 e_0^{24}-133 e_0^{12}+80}{14580e_0^8}\ \rho^2
+\calo(\rho^3)\,,
\eqlabel{zahrho}
\end{equation}
independent of $\mu$.

\begin{figure}[t]
\begin{center}
\psfrag{r}{{$\rho$}}
\psfrag{w}{{$\{A_v,{\color{blue} \sigma_v},{\color{red} Z_{AH}}\}$}}
\psfrag{s}{{$\{{\color{blue} z_v},{\color{magenta} \bar{z}_v},{\color{black} \eta_v}\}$}}
\includegraphics[width=2.6in]{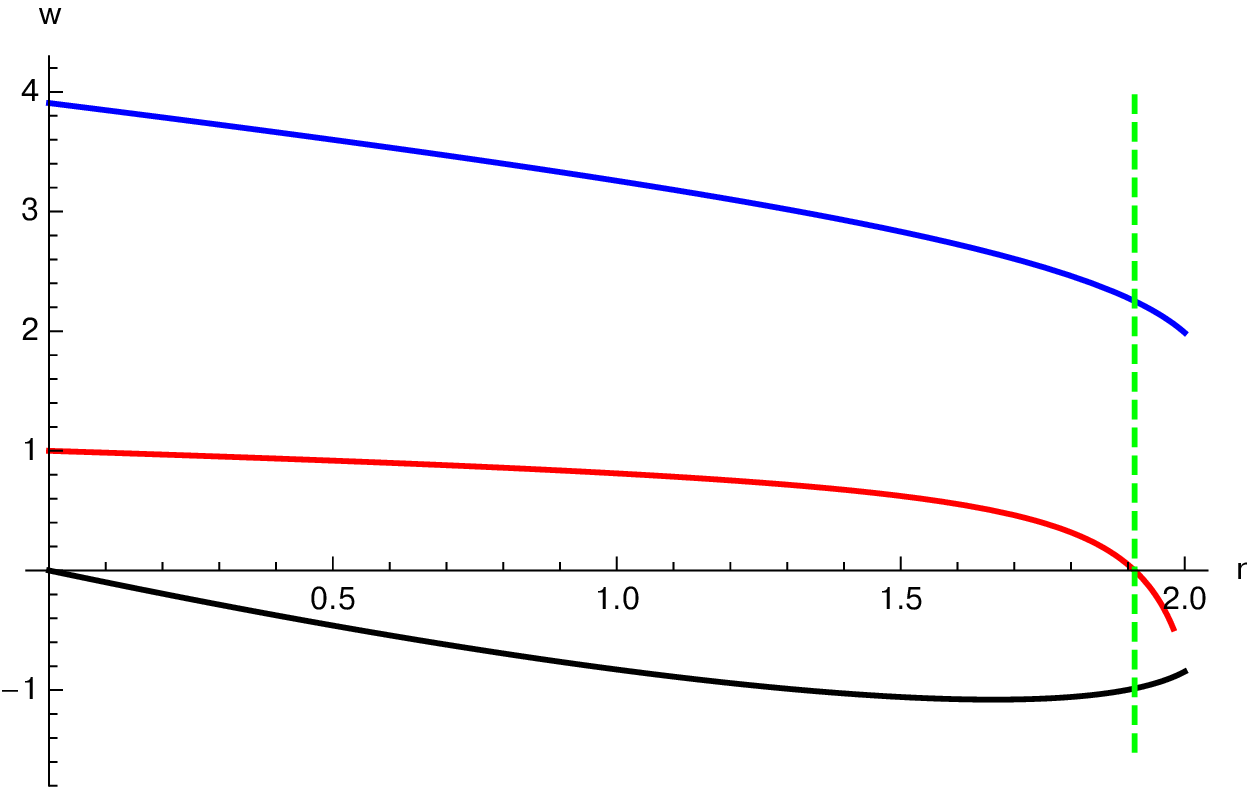}\qquad
\includegraphics[width=2.6in]{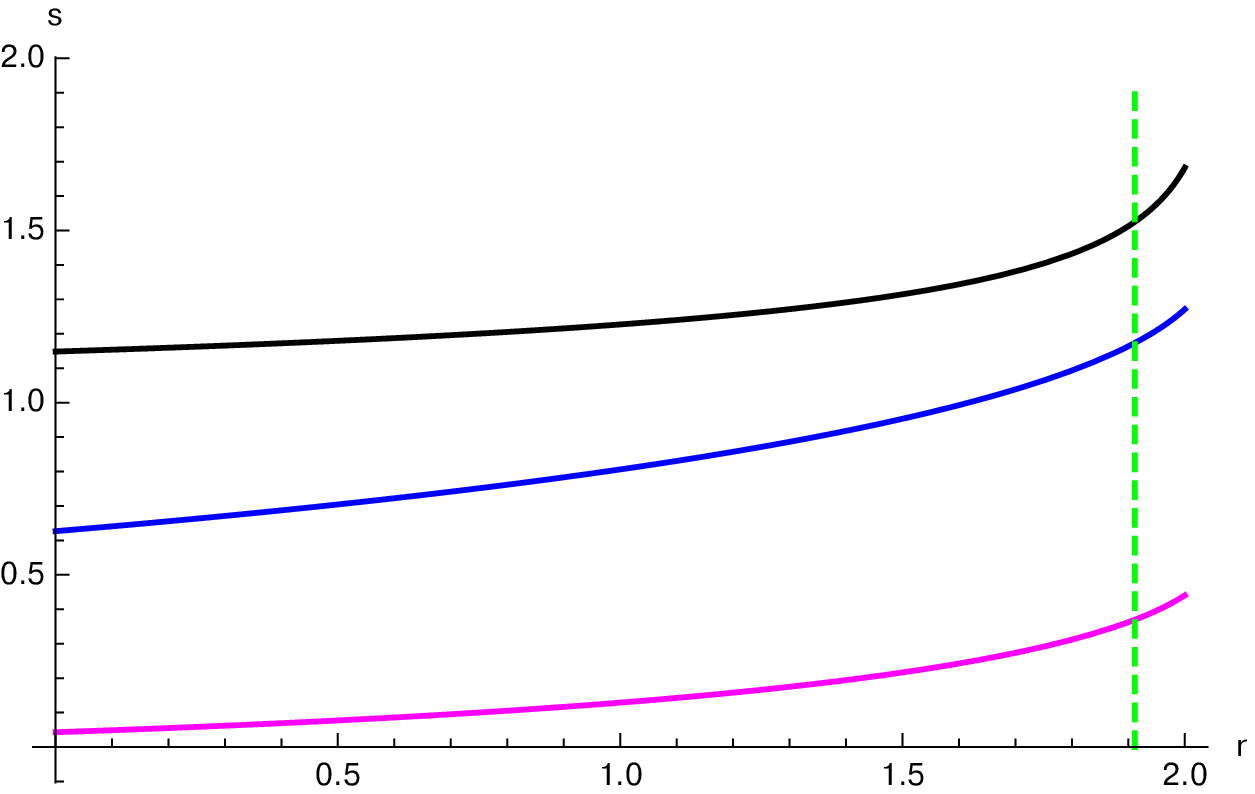}
\end{center}
  \caption{We set $\mu=0.9$. Left panel: profiles for $\{A_v,{\color{blue} \sigma_v},{\color{red} Z_{AH}}\}$. Right panel: profiles for $\{{\color{blue} z_v},{\color{magenta} \bar{z}_v},{\color{black} \eta_v}\}$. The vertical dashed green line indicates the location of the apparent horizon.
  The radial coordinate $\rho\in [0,\rho_{AH}+)$ \eqref{defrho} covers the second gravitational
  bulk subregion (see discussion in section \ref{split}).
 } \label{figure2}
\end{figure}

Typical profiles of the metric warp factors $\{A_v,\sigma_v\}$
(solid black and blue curves, left panel), the bulk scalars
$\{z_v,\bar{z}_v,\eta_v\}$ (solid blue, magenta and black curves, right panel),
and the AH monitoring function $Z_{AH}$  (solid red curve, left panel) are presented
in fig.~\ref{figure2} for $\mu=0.9$. The vertical dashed green line indicates the location of the
AH,
\begin{equation}
\rho_{AH}\bigg|_{\mu=0.9}=1.9117(4)\,.
\eqlabel{rhotyp}
\end{equation}
Most importantly, contrary to the $\caln=4$ SYM \eqref{vanishs}, we find here
\begin{equation}
\sigma_v\bigg|_{\rho=\rho_{AH}(\mu=0.9)}=2.2519(7)\ne 0\,,
\eqlabel{calrtyp1}
\end{equation}
resulting in
\begin{equation}
\begin{split}
&\calr_{(i\mu\,,\ \mu H/2)}^{\caln=2^*}=\frac{1}{H^3}\ s_{ent}= \frac{1}{4G_5}\ \sigma_v^3\bigg|_{\rho=\rho_{AH}}=
\frac{N^2}{16\pi}\ \sigma_v^3\bigg|_{\rho=\rho_{AH}}\qquad \Longrightarrow\qquad \\
&\frac{16\pi}{N^2}\ \calr_{(i\mu\,,\ \mu H/2)}^{\caln=2^*}\bigg|_{\mu=0.9}=11.420(6)\,.
\end{split}
\eqlabel{calrtyp2}
\end{equation}
We collect $\calr$ results for different values of $\mu$ at the end of this section.

\begin{figure}[t]
\begin{center}
\psfrag{x}{{$\mu$}}
\psfrag{y}{{$\hat{v}(\mu)$}}
\psfrag{z}{{$\dd \hat{v}(\mu)$}}
\includegraphics[width=2.6in]{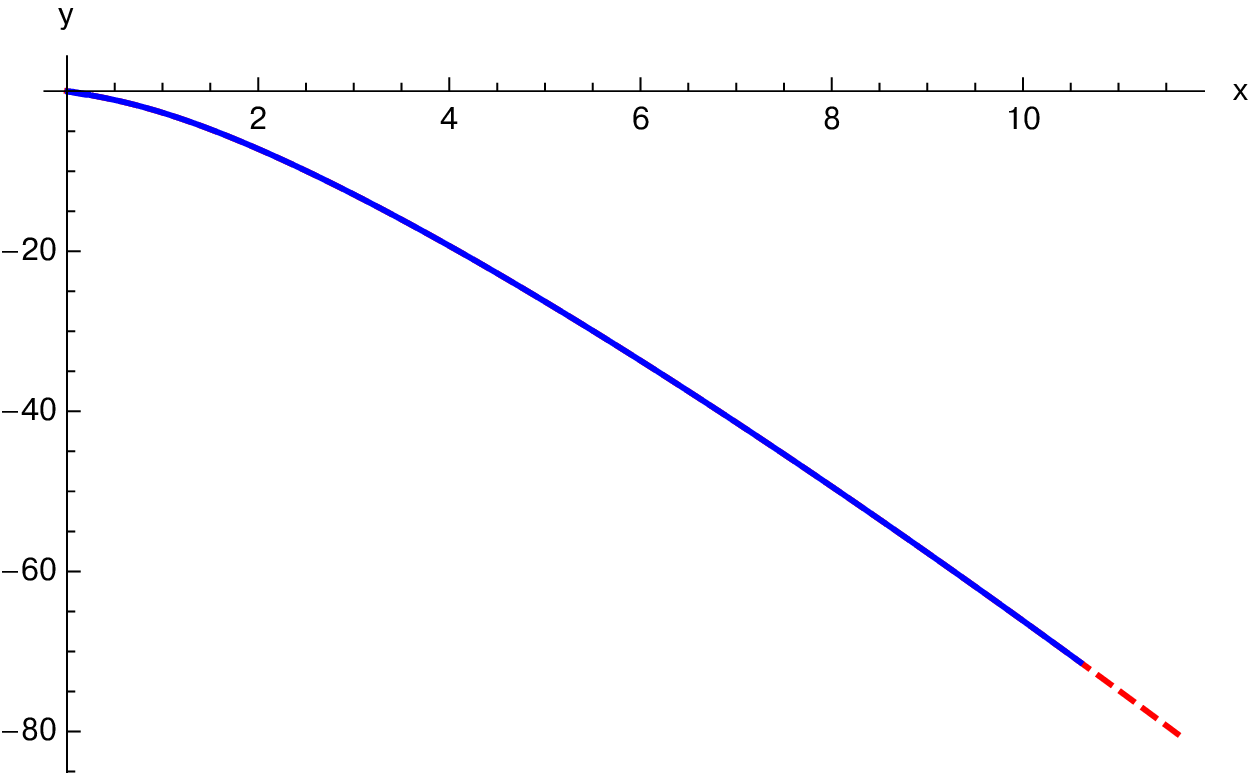}\qquad
\includegraphics[width=2.6in]{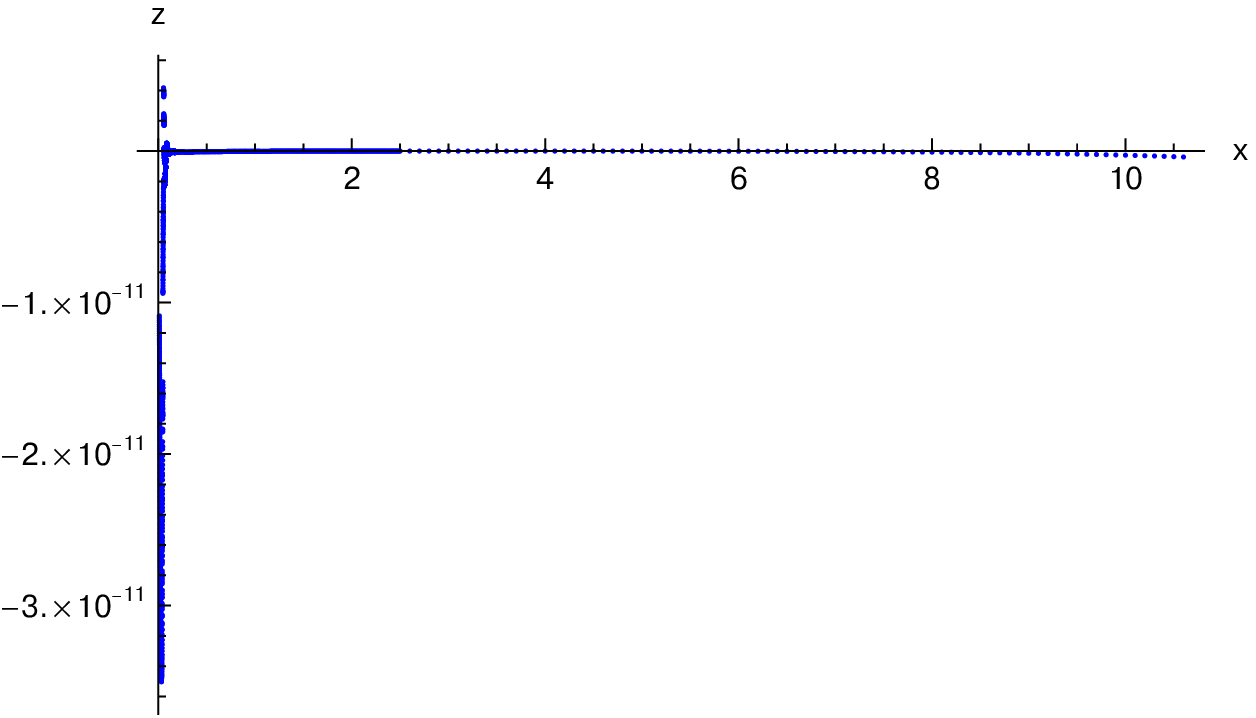}
\end{center}
  \caption{Left panel: numerical values for $\hat{v}(\mu)$ (see \eqref{uv21}) (solid blue curve)
  and the predicted values $\hat{v}_{prediction}$ (see \eqref{vpred2}) (dashed red curve). Right panel:
  the residual $\dd \hat{v}$ (see \eqref{defres2}).
 } \label{figure3}
\end{figure}

We now outline the differences in the second example considered
\begin{equation}
(m,k)=(\mu,i\mu/2)\,,\qquad \Im\mu=0\,.
\eqlabel{case2}
\end{equation}
All the analysis proceed literally unchanged once we introduce
\begin{equation}
z_v=i\ \hat{z}_v\,,\qquad \bar{z}_v=i\ \hat{\bar{z}}_v\,,
\eqlabel{changec2}
\end{equation}
so that \eqref{uv11} is modified to
\begin{equation}
\begin{split}
&\hat{z}_v=-\mu x+x^2 (2 \mu \ln x-\hat{v})+\calo(x^3\ln x)\,,
\\
&\hat{\bar{z}}_v=\mu x+x^2 (2 \mu \ln x-\hat{v})+\calo(x^3\ln x)\,.
\end{split}
\eqlabel{uv21}
\end{equation}
Eq.~\eqref{vpred} now reads
\begin{equation}
\hat{v}\bigg|_{prediction}=-2\mu-\mu \ln(1+\mu^2)\,.
\eqlabel{vpred2}
\end{equation}
Numerical results for $\hat{v}(\mu)$ and the residual $\dd\hat{v}$
\begin{equation}
\dd\hat{v}=1-\frac{\hat{v}(\mu)}{\hat{v}_{prediction}}\,,
\eqlabel{defres2}
\end{equation}
are collected in fig.~\ref{figure3}.

\begin{figure}[t]
\begin{center}
\psfrag{x}{{$\mu/H$}}
\psfrag{y}{{$\frac{16\pi}{N^2}\cdot \calr^{\caln=2^*}_{(i\mu,\mu H/2)}$}}
\psfrag{z}{{$\frac{16\pi}{N^2}\cdot \calr^{\caln=2^*}_{(\mu,i\mu H/2)}$}}
\includegraphics[width=2.6in]{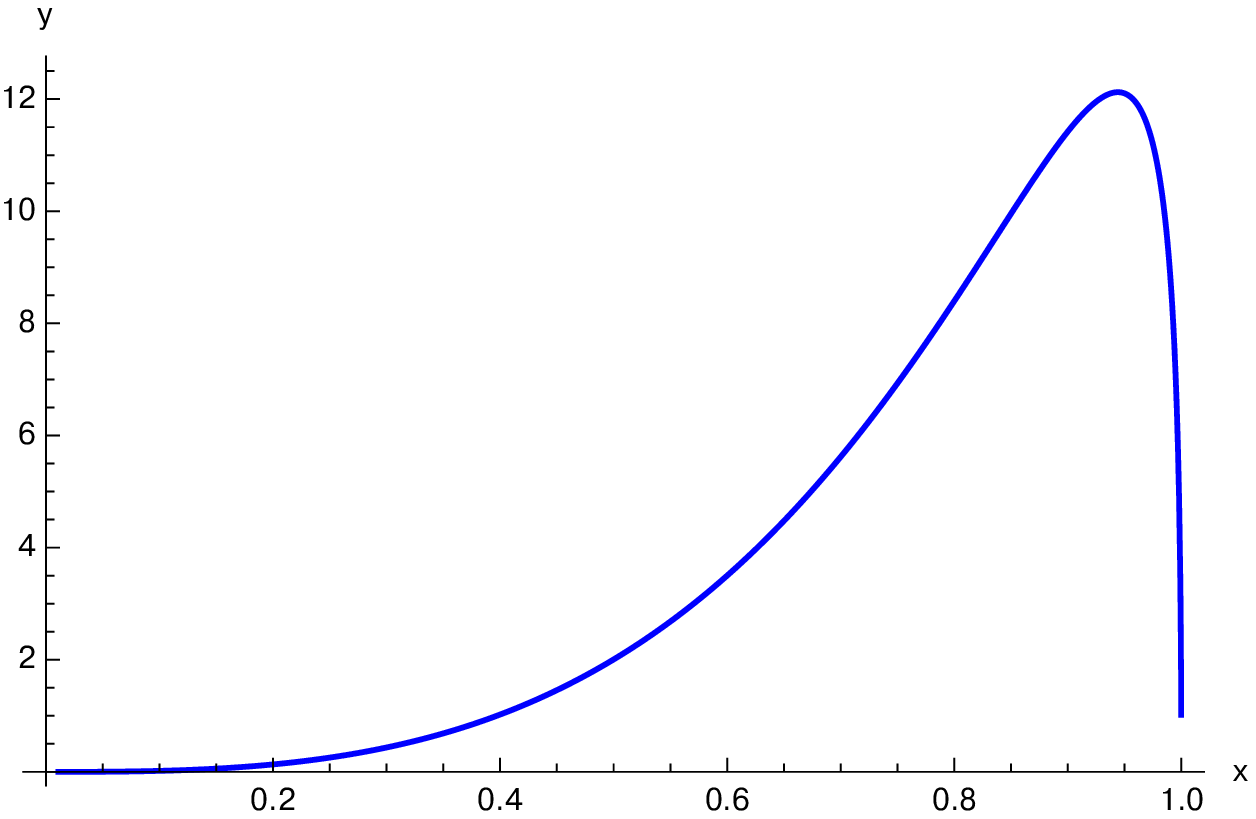}\qquad
\includegraphics[width=2.6in]{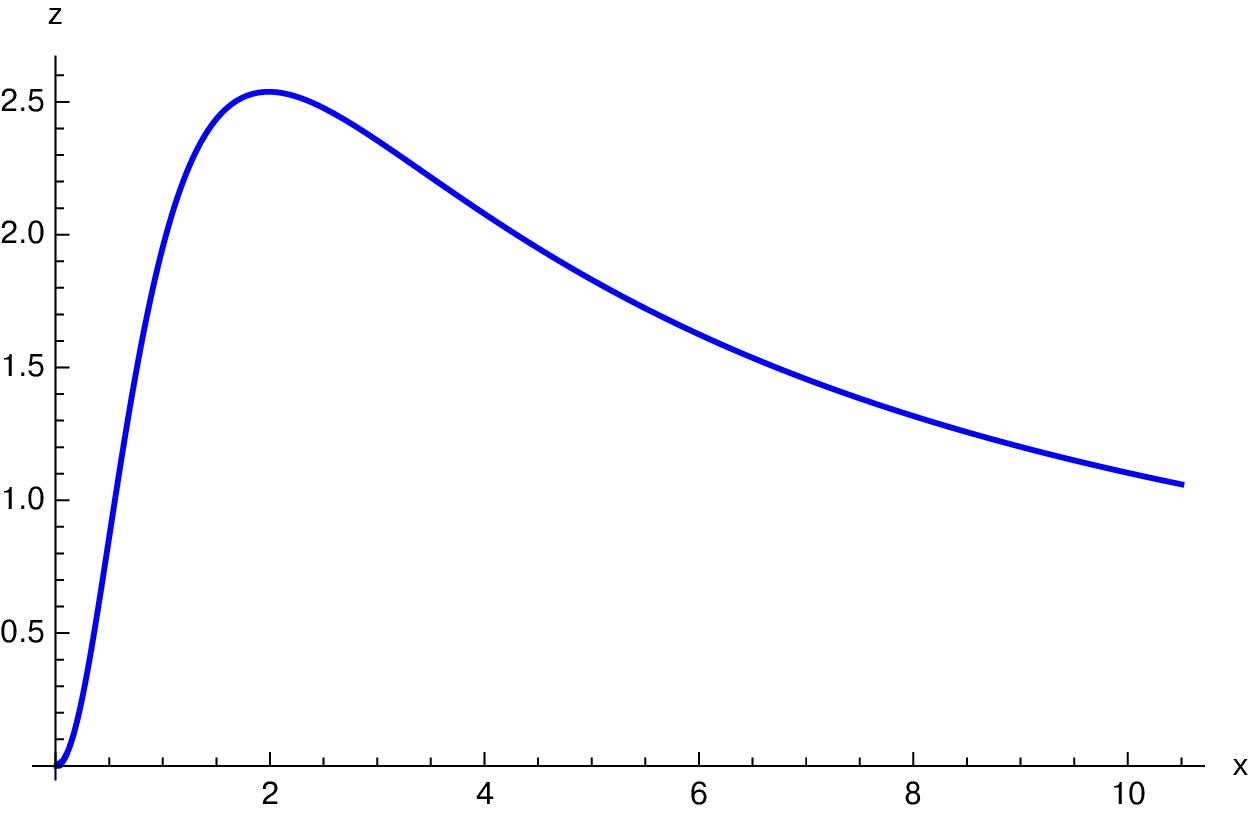}
\end{center}
  \caption{
$\calr^{\caln=2^*}_{(m=i\mu,k=\mu H/2)}$ (left panel) and $\calr^{\caln=2^*}_{(m=\mu,k=i\mu H/2)}$
(right panel) as a function of $\mu/H$. } \label{figure4}
\end{figure}

Fig.~\ref{figure4} is the main result of the section. It shows
$\calr^{\caln=2^*}_{(m=i\mu,k=\mu H/2)}$ (left panel) and $\calr^{\caln=2^*}_{(m=\mu,k=i\mu H/2)}$
(right panel) as a function of $\mu/H$. Unlike \eqref{vpred} and \eqref{vpred2},
we have been unable to guess the analytic expressions for $\calr$.
As we already stated in section \ref{intro}, the challenge for the localization is
to reproduce the predictions reported here.

\subsection{$\calr^{\caln=2^*}_{(m=\mu,k=\mu H/2)}$}\label{mainphysical}

With 
\begin{equation}
(m,k)=(\mu,\mu/2)\,,\qquad \Im\mu=0\,,
\eqlabel{case3}
\end{equation}
there is no (Euclidean) supersymmetry, and the holographic RG flows
are of the second order, see \eqref{veoms} and \eqref{veoms2}.
We discuss this case in some details, as the techniques developed are vital for
understanding de Sitter vacua of cascading gauge theories \cite{Klebanov:2000hb}. 

We use a new  radial coordinate $x\in (0,\infty)$ covering the first bulk subregion
(see section \ref{split})
\begin{equation}
r=\frac{H}{x}\,,
\eqlabel{eftofg3}
\end{equation}
implementing the transformation from EF \eqref{efbmetric} to FG coordinate system
\begin{equation}
\begin{split}
&ds_5^2=\frac{H^2}{x^2 h^{1/2}(x)}\left(-(d\tau)^2+\frac{1}{H^2}\cosh^2(H\tau)\left(dS^3\right)^2\right)
+\frac{h^{1/2}(x)}{x^2}\ dx^2\,,\\
&d\tau =dt+\frac{h^{1/2}(x)}{H}\ dx\,, 
\end{split}
\eqlabel{defg3}
\end{equation}
where we impose
\begin{equation}
\lim_{x\to \infty} x^2 h(x)=\frac 14\,,
\eqlabel{limh}
\end{equation}
which ensures that the geometry \eqref{defg3} is smooth in the limit $y\equiv \frac 1x\to 0$
\begin{equation}
\begin{split}
&ds_5^2\sim \left(\sqrt{2y}\right)^2\ \left(-(Hd\tau)^2+\cosh^2(H\tau)\left(dS^3\right)^2\right)+\left(d\sqrt{2y}\right)^2
+\calo(y^2)\,,\\
&ds_{5,E}^2\underbrace{\sim}_{\tau\to \tau_E=i\theta/H} \left(\sqrt{2y}\right)^2\ \left( dS^4\right)^2+\left(d\sqrt{2y}\right)^2
+\calo(y^2)\,.
\end{split}
\eqlabel{smallylim}
\end{equation}
The full set of the vacuum equations of motion obtained from \eqref{veoms} and \eqref{veoms2} is collected in
appendix \ref{physmass}. Note that we introduced
\begin{equation}
\sigma(x)=\frac{H s(x)}{x h^{1/4}(x)}\,,\qquad s(y\equiv\frac 1x)= \frac{\hat{s}(y)}{y^{1/2}}\,.
\eqlabel{defsigmac3}
\end{equation}
Equations \eqref{case3r1} must be solved numerically subject to the following asymptotes:
\begin{itemize}
\item asymptotic $AdS$ boundary, \ie $x\to 0_+$,
\begin{equation}
\begin{split}
&z_{1,v}=x^2\left(z_{1,2,0}+8\mu\k \ln x\right)+\calo(x^3\ln x)\,,\\
&z_{2,v}=2\mu x +\frac{\mu h_1}{32} x^2 +x^3\left(z_{2,3,0}+\left(\frac{32}{3}\mu^3+32\mu\right)\ln x\right)
+\calo(x^4\ln x)\,,\\
&\eta_v=1+x^2\left(e_{2,0}+\frac 83\mu^2\ln x\right)+\calo(x^3\ln x)\,,\\
&h=16+h_1 x +\calo(x^2)\,,\qquad s=1+4x+\calo(x^2)\,,
\end{split}
\eqlabel{uv31}
\end{equation}
where we showed explicitly the dependence on normalizable coefficients $\{z_{1,2,0},z_{2,3,0},h_1,e_{2,0}\}$;
the parameter\footnote{There is an exact $\zet_2$ symmetry $z_{1,v}\leftrightarrow -z_{1,v}$ implying that
 the choice of a sign of $\k$ is physically irrelevant.} $\k=\{\pm 1,0\}$ corresponding to $k=\k\mu/2$;
\item $h\to 0$, \ie $y\equiv \frac 1x\to 0_+$,
\begin{equation}
\begin{split}
&z_{1,v}=z_{1,0}^h+\calo(y)\,,\qquad z_{2,v}=z_{2,0}^h+\calo(y)\,,\qquad \eta_v=\eta^h_0+\calo(y)\,,\\
&h=\frac 14 y^2 +\calo(y^3)\,,\qquad \hat{s}=s^h_0+\calo(y)\,.
\end{split}
\eqlabel{ir31}
\end{equation}
\end{itemize}

Notice that fixing $\{\mu,\k\}$, the holographic RG flows are completely determined by
8 parameters
\begin{equation}
\{z_{1,2,0},z_{2,3,0},h_1,e_{2,0}\}\ \bigcup\ \{z_{1,0}^h,z_{2,0}^h,\eta_0^h,s_0^h\}\,,
\eqlabel{par8}
\end{equation}
which is the correct overall number to  specify a solution for a system \eqref{case3r1} of
3 second-order differential equations and 2 first-order differential equations.
This has to be contrasted with the (Euclidean) supersymmetric flows \eqref{case1r1},
which are determined by 4 parameters: $\{v,x^*,e_0,s_0\}$ (see \eqref{uv11} and \eqref{ir11}).
Of course, these supersymmetric flows represent a special case of the RG flows discussed in this section:
after relating the radial coordinates, the UV parameters \eqref{uv31} and \eqref{uv11} are matched as follows
\begin{center}
 \begin{tabular}{||c | c||} 
 \hline
 Eq.~\eqref{uv31} & Eq.~\eqref{uv11}  \\ [0.5ex] 
 \hline\hline
 $\mu$ & $i \mu$  \\ 
 \hline
 $z_{1,2,0}$ & $4v-8\mu\ln 2$  \\ 
 \hline
 $i z_{2,3,0}$ & $\left(\frac{32}{3}\ln 2-8\right)\mu^3-\frac{16}{3}\mu^2 v
 -\left(\frac{h_1^2}{2048}+32\ln 2\right)\mu+16v$  \\ 
 \hline
  $e_{2,0}$ & $\frac43 v\mu+\left(\frac 43-\frac83\ln 2\right)\mu^2$  \\ 
 \hline
   \end{tabular}
\end{center}
Parameter $h_1$ remains unfixed in UV; this is related to the residual symmetry of the
metric ansatz \eqref{eftofg3}
\begin{equation}
x\to \frac{x}{1+\a x}\,,\qquad h\to (1+\a x)^4 h\,,\qquad \a={\rm const}\,.
\eqlabel{resxtra}
\end{equation}
This symmetry is fixed with the boundary condition in IR, see \eqref{limh}.

We use the numerical shooting method adopted from \cite{Buchel:2013dla} to construct RG flows
for various values of $\mu$ and $\k=\{1,0\}$ realizing the first subregion of the dual geometry as per discussion in
section \ref{split}. To compute the comoving entropy production rate $\calr^{\caln=2^*}_{(\mu,\k\mu/2)}$
we need an access to the second subregion of the bulk geometry with $A_v\le 0$.
This is done returning to the original ER coordinate
$r$, see \eqref{eftofg1}. Introduce
\begin{equation}
r=-\rho\equiv y\,,\qquad \rho\in [0,\rho_{AH}]\ \Longleftrightarrow\ r\in[r_{AH,vacuum},0]\,.
\eqlabel{derrho3}
\end{equation}
The holographic equations of motion in $\rho$ coordinate are simply \eqref{veoms} with
\[
\del_r\ \equiv\ '\ =\ -\del_\rho\,.
\]
Given \eqref{ir31} and the identification $\rho\equiv -y$, it is trivial to provide asymptotic initial conditions
for  \eqref{veoms} (eqs.~\eqref{veoms2} are satisfied as well, but we will not use them)
\begin{equation}
\begin{split}
&A_v\equiv \frac{y^2}{2h^{1/2}(y)}\bigg|_{y=-\rho}
=-\rho-\frac{1}{24}\biggl(
\frac{(\eta^h_0)^8 (z_{2,0}^h)^2}{((z_{1,0}^h)^2+(z_{2,0}^h)^2-1)^2}
+\frac{2 ((z_{1,0}^h)^2+(z_{2,0}^h)^2+1) (\eta^h_0)^2}
{(z_{1,0}^h)^2+(z_{2,0}^h)^2-1}\\
&\qquad\qquad\qquad\qquad\qquad  -\frac{1}{(\eta^h_0)^4}
\biggr) \rho^2+\calo(\rho^3)\,,\\
&\sigma_v=\frac{y^{1/2}\hat{s}(y)}{h^{1/4}(y)}\bigg|_{y=-\rho}
=s^h_0 \sqrt{2} \biggl(
1+\frac{1}{24} \biggl(\frac{(\eta^h_0)^8 (z_{2,0}^h)^2}
{((z_{1,0}^h)^2+(z_{2,0}^h)^2-1)^2}\\
&\qquad\qquad\qquad\qquad\qquad\qquad\qquad  +\frac{2 ((z_{1,0}^h)^2+(z_{2,0}^h)^2+1)
(\eta^h_0)^2}{(z_{1,0}^h)^2+(z_{2,0}^h)^2-1}-\frac{1}{(\eta^h_0)^4}
\biggr) \rho\biggr)+\calo(\rho^2)\,,\\
&z_{1,v}=z_{1,0}^h+\frac{z_{1,0}^h(\eta^h_0)^2}{10}  \left(
\frac{(\eta^h_0)^6 (z_{2,0}^h)^2}{(z_{1,0}^h)^2+(z_{2,0}^h)^2-1}+2\right) \rho+\calo(\rho^2)\,,\\
&z_{2,v}=z_{2,0}^h-\frac{z_{2,0}^h (\eta^h_0)^2}{20}
\left(\frac{(\eta^h_0)^6 ((z_{1,0}^h)^2-(z_{2,0}^h)^2-1)}
{(z_{1,0}^h)^2+(z_{2,0}^h)^2-1}-4\right) \rho+\calo(\rho^2)\,,\\
&\eta_v=\eta^h_0-\frac{1}{30} \biggl(
\frac{2 (\eta^h_0)^9 (z_{2,0}^h)^2}{((z_{1,0}^h)^2+(z_{2,0}^h)^2-1)^2}
+\frac{((z_{1,0}^h)^2+(z_{2,0}^h)^2+1) (\eta^h_0)^3}
{(z_{1,0}^h)^2+(z_{2,0}^h)^2-1}+\frac{1}{(\eta^h_0)^3}
\biggr) \rho+\calo(\rho^2)\,.
\end{split}
\eqlabel{sing3}
\end{equation}
AH-monitoring function here is
\begin{equation}
Z_{AH}\equiv \sigma_v-A_v\frac{d\sigma_v}{d\rho}\,,\qquad Z_{AH}\bigg|_{\rho=\rho_{AH}}=0\,.
\eqlabel{zah3}
\end{equation}

\begin{figure}[t]
\begin{center}
\psfrag{x}{{$\mu^2/H^2$}}
\psfrag{y}{{$\frac{16\pi}{N^2}\cdot \calr^{\caln=2^*}_{(\mu,\k\mu H/2)}$}}
\psfrag{z}{{$\frac{16\pi}{N^2}\cdot \calr^{\caln=2^*}_{(\mu,0)}$}}
\includegraphics[width=2.6in]{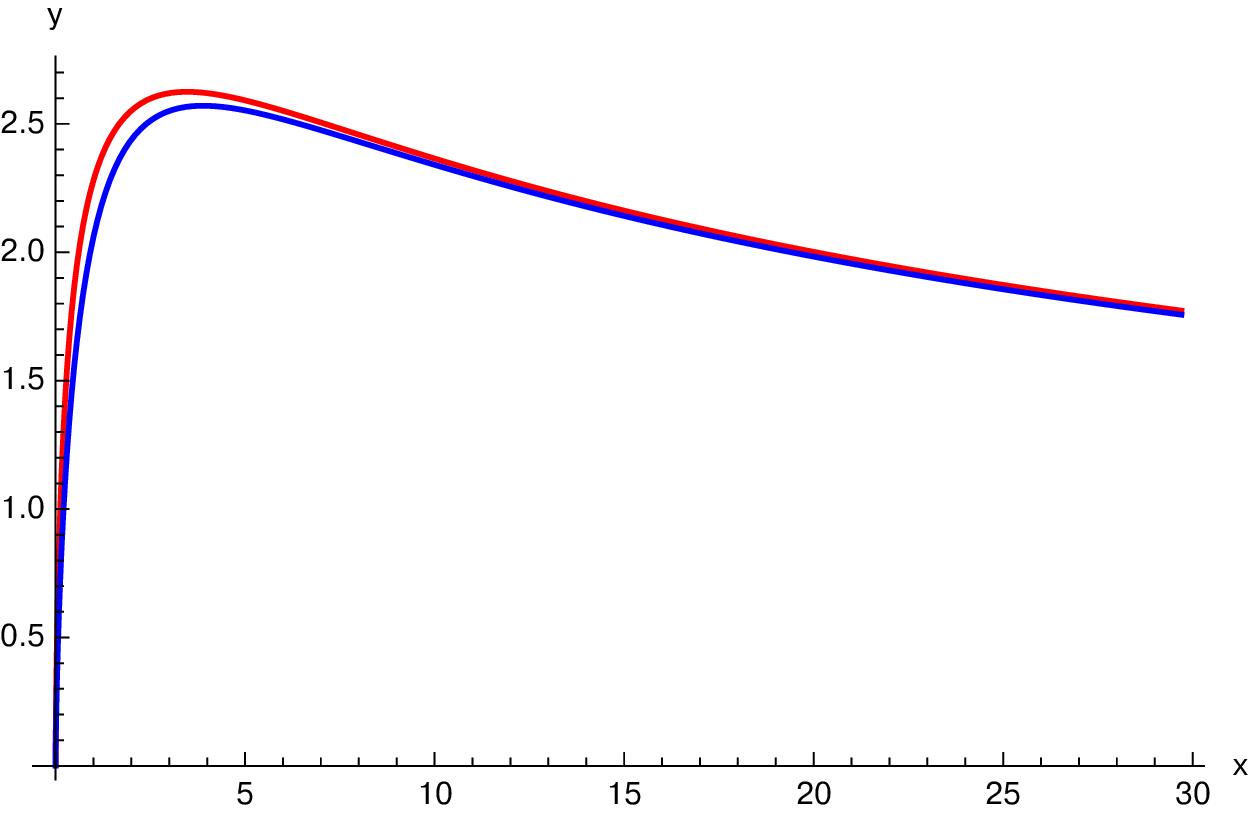}\qquad\qquad
\includegraphics[width=2.6in]{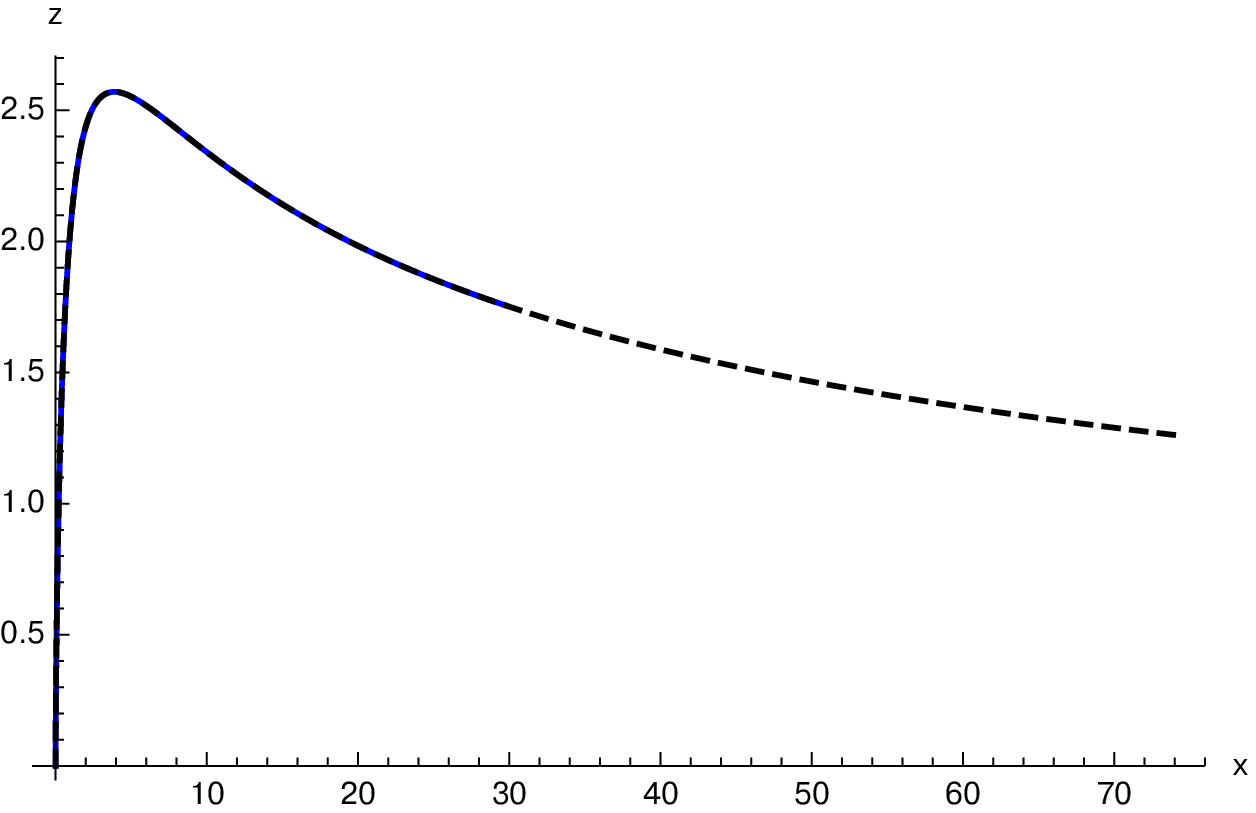}
\end{center}
  \caption{
$\calr^{\caln=2^*}_{(m=\mu,k=\k\mu H/2)}$ for $\color{red} \k=1$  and
$\color{blue} \k=0$ (left panel)  as a function of $\mu^2/H^2$. Right panel shows comparison
of the results for $\calr^{\caln=2^*}_{m=\mu,k=0}$ obtained using the
method discusses  here (solid blue curve) and the
equivalent results obtained in \cite{Buchel:2017pto} (dashed black curve).} \label{figure5}
\end{figure}

Fig.~\ref{figure5} is the main result of the section. It shows
$\calr^{\caln=2^*}_{(m=\mu,k=\k\mu H/2)}$ for $k=\{1,0\}$ (left panel, $\{$red,blue$\}$ curves).
The effect of the curvature coupling $k$ is relatively small and decreases as $\mu$ get larger ---
this is easy to understand, given that $\frac{k}{\mu^2}\sim \frac{H}{\mu}\to 0$ as $\frac{\mu^2}{H^2}\to \infty$.
Recall that the computations in this section were performed splitting the
holographic geometry into two subregions (see discussion in section \ref{split}):
from the asymptotic $AdS$ boundary to $A_v=0$ (the first subregion) and from
$A_v=0$ to the apparent horizon (the second subregion). We used Fefferman-Graham (Schwarzschild)
coordinates in the first subregion and  Eddington-Finkelstein coordinates in the second subregion.
The transition between the two coordinate systems was implemented at $A_v=0$. On the contrary, the computations
 of $\calr^{\caln=2^*}_{m=\mu,k=0}$ in \cite{Buchel:2017pto} were performed entirely in  Eddington-Finkelstein coordinate
 system. Comparison of the two approaches is shown in right panel, fig.~\ref{figure5}: blue solid curve
 is the new result, dashed black curve represent results obtained in \cite{Buchel:2017pto}.
 The agreement is $\sim 10^{-4}$ for $\mu < H$ and $\sim 10^{-6}$ (and rapidly improving) for $\mu>H$.
 This validates the computational method for the comoving entropy production rate $\calr$ developed here.

\section{Discussion}\label{discussion}

This paper is a continuation of the exploration of de Sitter vacua of non-conformal gauge theories
initiated in  \cite{Buchel:2017pto}.

Thermal equilibrium states of interacting QFTs in Minkowski space-time are
(typically\footnote{There are exceptions \cite{Balasubramanian:2014cja}.}) universal end-points of late-time
dynamical evolution. Likewise, de Sitter vacua of interactive QFTs are also universal end-points of
late-time dynamical evolution  \cite{Buchel:2017pto}. However, these vacua are definitely not the equilibrium states.
There are simple ways to see this:
\nxt first, it is inconsistent to recast late-time evolution in de Sitter as hydrodynamics --- the ``local temperature''
dilutes with the metric scale factor as $e^{- H t}$, while the typical velocity gradients remain constant $|\del u| \sim H$,
and thus the standard hydrodynamic approximation, \ie $|\del u|/T_{loc}\ll 1$ is not valid;
\nxt second, there is no comoving entropy production in equilibrium states; here, unless the quantum field theory is conformal,
there is non-vanishing comoving entropy production rate $\calr$ at late time $ H t \gg 1$.

The comoving entropy production rate $\calr$ in de Sitter vacuum implies that there is a nonzero vacuum entropy density
\begin{equation}
s_{ent}=H^3\ \calr\,.
\eqlabel{sdis}
\end{equation}
The latter quantity is renormalization scheme independent, and might serve as a valuable tool to classify symmetry breaking
phases of the theories in de Sitter\footnote{We discuss this in forthcoming publication.}.

\begin{figure}[t]
\begin{center}
\psfrag{x}{{$\mu^2/H^2$}}
\psfrag{y}{{$\frac{16\pi}{N^2}\cdot \calr^{\caln=2^*}_{(0,\mu H/2)}$}}
\includegraphics[width=4in]{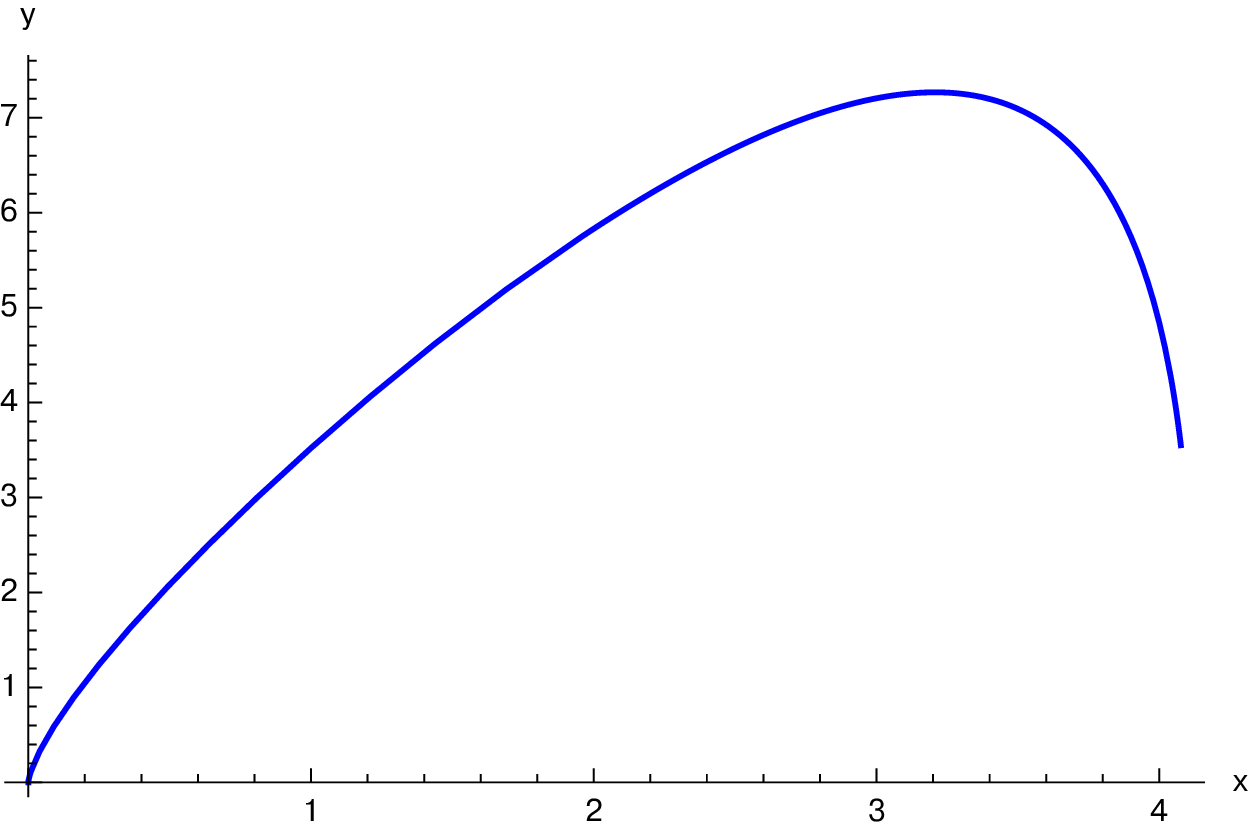}
\end{center}
  \caption{
$\calr^{\caln=2^*}_{(m=0,k=\mu H/2)}$  as a function of $\mu^2/H^2$.
Here, the conformal symmetry is broken by the nonvanishing coupling $k$ to the
background space-time, see \eqref{deflambda}. The profile for the comoving entropy
production rate is similar to the one in $\caln=2^*$ gauge theory with $k=0$, and different
masses for the bosonic $m_b\ne 0$ and fermionic $m_f=0$ components  in $\caln=2$ hypermultiplet,
see left panel of fig.~8 in \cite{Buchel:2017pto}.
}\label{figure6}
\end{figure}

In this paper we expended the computations of $\calr$ in $\caln=2^*$ gauge theory in  \cite{Buchel:2017pto},
including the curvature coupling $k$ (see \eqref{deflambda}). We showed that unless both $\mu=0$ and $k=0$
(see fig.~\ref{figure6} for results for $\calr^{\caln=2^*}_{(m=0,k=\mu H/2)}$; this is a special case of the general approach
considered in section \ref{mainphysical}) the comoving entropy production rate is non-zero. We developed a new approach
towards computation of $\calr$: the dual gravitational geometry is split into two subregions --- from the
asymptotic $AdS$ boundary to $g_{tt}=0$, and from $g_{tt}=0$ to the apparent horizon ---   with Fefferman-Graham (Schwarzschild)
coordinates used in the first subregion, and  the Eddington-Finkelstein coordinates used in the second subregion.
The transition between the two subregions is implemented at $g_{tt}=0$. This approach differs from the one employed in
 \cite{Buchel:2017pto}, where the Eddington-Finkelstein coordinates covered the whole bulk geometry.
 We showed that both methods produce identical results, whenever appropriate. The advantage of the
 newly proposed method in that it allows the computation of $\calr$ in the theories where it is
 difficult (impossible?) to define the asymptotic Eddington-Finkelstein coordinates (as it is in the case of
 cascading gauge theories \cite{Klebanov:2000hb}).

For generic values of $(m,k)$
there is no supersymmetry. However, when 
\begin{equation}
k=\pm i\ \frac{Hm}{2}\,,
\eqlabel{ksusy}
\end{equation}
the holographic RG flow equations are of the first order. The BPS-like character of the flow equations
reflects the fact that the Wick rotation $H \tau\to i\theta$ of the first bulk subregion \eqref{deffg1}
represents supersymmetric holographic dual to $\caln=2^*$ gauge theory on $S^4$ \cite{Pestun:2007rz,Bobev:2013cja,Bobev:2018hbq}.
Powerful techniques of supersymmetric localization exist to compute plethora of properties of strongly coupled
$\caln=2^*$ gauge theory without resorting to holography
\cite{Pestun:2007rz,Buchel:2013id,Chen:2014vka,Zarembo:2014ooa,Chen-Lin:2017pay,Russo:2019lgq}. 
We computed $\calr^{\caln=2^*}_{(m,k=im H/2)}$ in section \ref{mainmuimu}. We hope that the supersymmetric
localization of \cite{Pestun:2007rz} will ultimately shed light on this quantity and its physical origin.

We conclude with our two failed attempts to interpret $s_{ent}$.

\subsection{$s_{ent}$ as a thermal entropy of pair-produced particles}

In a dynamical expanding Universe 
a minimally coupled free massless scalar field experiences particle production.  
While particle pair production in different momentum modes are independent events,
pairs produced in a given mode correlate. Local observers will be unable
to detect the correlations for produced pairs separated by cosmological scales.
As a result, the observed spectrum of produced particles is
thermal\footnote{See \cite{Parker:2009uva} for a nice presentation.}.

\begin{figure}[t]
\begin{center}
\psfrag{x}{{$m^2/T_{dS}^2$}}
\psfrag{y}{{$\frac{16\pi}{N^2}\cdot \frac{s}{T_{dS}^3}$}}
\includegraphics[width=4in]{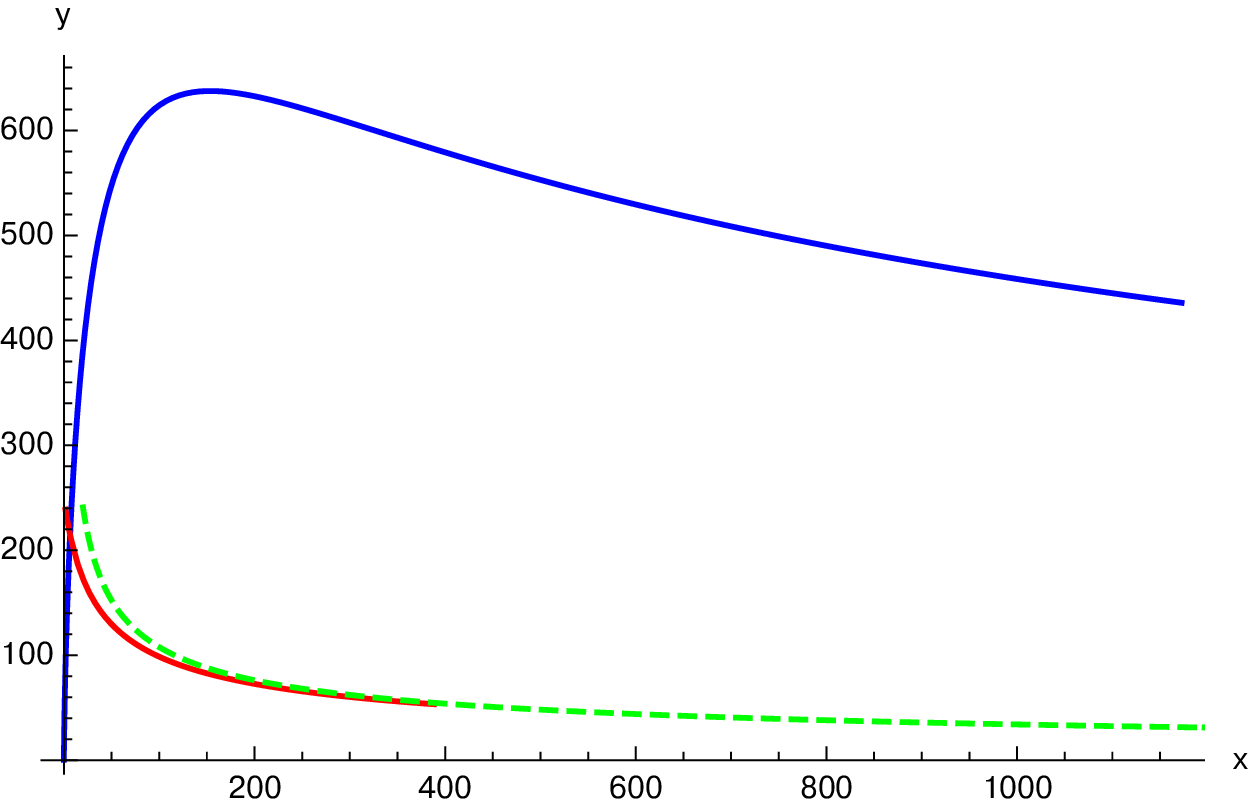}
\end{center}
  \caption{
de Sitter vacuum entanglement entropy density $\color{blue} s_{ent}$ of $\caln=2^*$ gauge theory
at $(m\ne 0, k=0)$ (solid blue curve) and the thermal entropy density of the theory $\color{red} s_{thermal}$
as a function of $m^2/T^2$ with $T=T_{dS}=H/(2\pi)$. Green dashed curve is the analytic asymptote of
$\color{green} s_{thermal}$ as  $m^2/T^2\to \infty$, see \eqref{greenass}. 
}\label{figure7}
\end{figure}

For an interactive quantum field theory in de Sitter space-time one expects that the pair production
will be incoherent at late times, leading to a thermal spectrum with de Sitter temperature $T_{dS}$ \eqref{tempdS}
(see also \cite{Spradlin:2001pw}). Thus, it is tempting to identify
\begin{equation}
s_{ent}=s_{thermal}\bigg|_{T=T_{dS}=\frac{H}{2\pi}}\,.
\eqlabel{theds}
\end{equation}
Unfortunately, the identification \eqref{theds} can not be correct:
\nxt it is incorrect for conformal field theories, where $s_{ent}^{CFT}\propto \calr^{CFT}=0$, while
$s_{thermal}^{CFT}\propto H^3$;
\nxt as fig.~\ref{figure7} shows, it is also incorrect for $\caln=2^*$ gauge theory. 

Fig.~\ref{figure7} presents results for the entanglement entropy $s_{ent}$ for $\caln=2^*$ gauge theory with $(m\ne 0,k=0)$
as a function of $m^2/T_{dS}^2$ (the solid blue curve), and the thermodynamic entropy of the theory at equivalent
value $T=T_{dS}$ (the solid red curve).
The dashed green curve represents the asymptotic of the thermodynamic entropy as $m^2\gg T_{dS}^2$,
\begin{equation}
\frac{16\pi}{N^2}\cdot \frac{s_{thermal}}{T_{dS}^3}\ \sim\ \frac{6912\pi^4}{625} \left(\frac{m^2}{T_{dS}^2}\right)^{-1/2}\,,
\eqlabel{greenass}
\end{equation}
computed in appendix \ref{IRthermodynamics}.
Note that
\begin{equation}
\frac{16\pi}{N^2}\cdot \frac{s_{thermal}}{T_{dS}^3}\bigg|_{m^2/T_{dS}^2\to 0}=8\pi^3\,.
\eqlabel{st3cft}
\end{equation}
From fig.~\ref{figure7} it is clear that there is no obvious relation between $s_{ent}$
and a "subtracted'' thermodynamic entropy
\begin{equation}
\frac{s_{subtracted}}{T^3}\equiv \frac{s_{thermal}}{T^3}\bigg|_{T=T_{dS}}-\frac{s_{thermal}}{T^3}\bigg|_{T\to \infty}
\end{equation}
--- at the very least, by definition, $s_{subtracted}$ vanishes for CFTs.

\subsection{$s_{ent}$ as a thermodynamic entropy of the localization free energy at $T=T_{dS}$}

In holographic thermodynamics, the thermal free energy density $f_{thermal}$ is related to the renormalized Euclidean bulk
gravitational action $\cali_E$ as follows
\begin{equation}
\cali_E=\underbrace{\int_{\calm_3}d^3x \int dt_E}_{\int_{\calm_4^E}d\xi^4}\ f_{thermal}\,,
\eqlabel{fi}
\end{equation}
where $\int_{\calm_3}d^3x$ is the spatial integral, and $\int dt_E=\frac 1T$ is the integral over the (compactified) Euclidean
temporal direction. We denoted $\int_{\calm_4^E}d\xi^4$ as the integral over the Euclidean space-time.
From \eqref{fi}, the average free energy density
\begin{equation}
\langle f_{thermal}\rangle= \frac{1}{{\rm vol}\calm_4^E}\ \cali_E\,.
\eqlabel{avf}
\end{equation}

We would like to apply \eqref{avf} to $\cali_E=\calf$, with $\calf$ computed either via supersymmetric localization \cite{Buchel:2013id}
or in holography \cite{Bobev:2013cja} (up to scheme dependence, both computations agree).
Using
\begin{equation}
{\rm vol}\calm_4^E=\frac{8}{3}\pi^2\ \ell_{S^4}^4=\frac{8\pi^2}{3H^4}\,,
\eqlabel{volm4}
\end{equation}
where the radius of $S^4$ is $\ell_{S^{4}}=1/H$, we find
\begin{equation}
\frac{16\pi}{N^2} \langle f_{thermal}\rangle=-\frac{3H^4}{\pi}\left(\left(1+\frac{m^2}{H^2}\right)
\ln\left(1+\frac{m^2}{H^2}\right)+\a_0+\a_1\ \frac mH+\a_2\ \frac{m^2}{H^2}\right)\,;
\eqlabel{fthres1}
\end{equation}
the arbitrary coefficients $\a_i$ parameterize the {\it full} renormalization scheme dependence
\cite{Buchel:2013fpa}.
Applying the first law of thermodynamics to $\langle f_{thermal}\rangle$ with $H=2\pi T_{dS}$, we identify
\begin{equation}
s_{ent,loc}=-\frac{d\langle f_{thermal}\rangle}{d T_{dS}}\,,
\eqlabel{sloc}
\end{equation}
leading to (see \eqref{s})
\begin{equation}
\begin{split}
\frac{16\pi}{N^2}\cdot \calr_{loc}\equiv& \frac{16\pi}{N^2}\cdot \frac{s_{ent,loc}}{(2\pi T_{dS})^3}\equiv \frac{16\pi}{N^2}\cdot \calr_{loc}^{0}+\frac{16\pi}{N^2}\cdot \calr_{loc}^{ambiguity}\\
=&12\left(2+\frac{m^2}{H^2}\right)
\ln\left(1+\frac{m^2}{H^2}\right)-\frac{12m^2}{H^2}+\frac{16\pi}{N^2}\cdot \calr_{loc}^{ambiguity}\,,
\end{split}
\eqlabel{rloc}
\end{equation}
where
\begin{equation}
\frac{16\pi}{N^2}\cdot \calr_{loc}^{ambiguity}=24\a_0+18\a_1\ \frac{m}{H}+12\a_2\ \frac{m^2}{H^2}\,,
\eqlabel{rscheme}
\end{equation}
parameterizes the scheme dependence. If we require the correct CFT limit (see \eqref{cftlimit}),
we must set $\a_0$. The ambiguity then is in the choice $\{\a_1,\a_2\}$.

\begin{figure}[t]
\begin{center}
\psfrag{x}{{$\mu/H$}}
\psfrag{y}{{$\frac{16\pi}{N^2}\cdot \dd\calr$}}
\psfrag{z}{{$\frac{16\pi}{N^2}\cdot (\dd\calr-\calr_{loc}^{ambiguity})$}}
\includegraphics[width=2.6in]{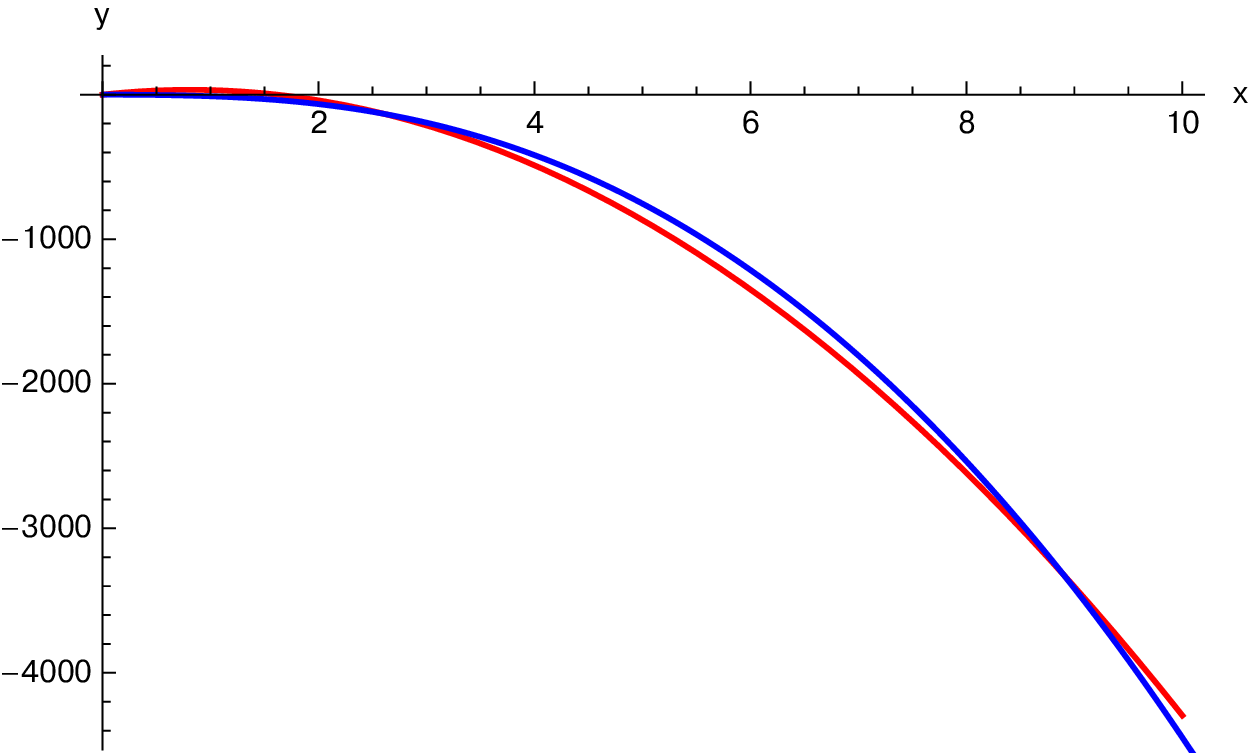}\qquad
\includegraphics[width=2.6in]{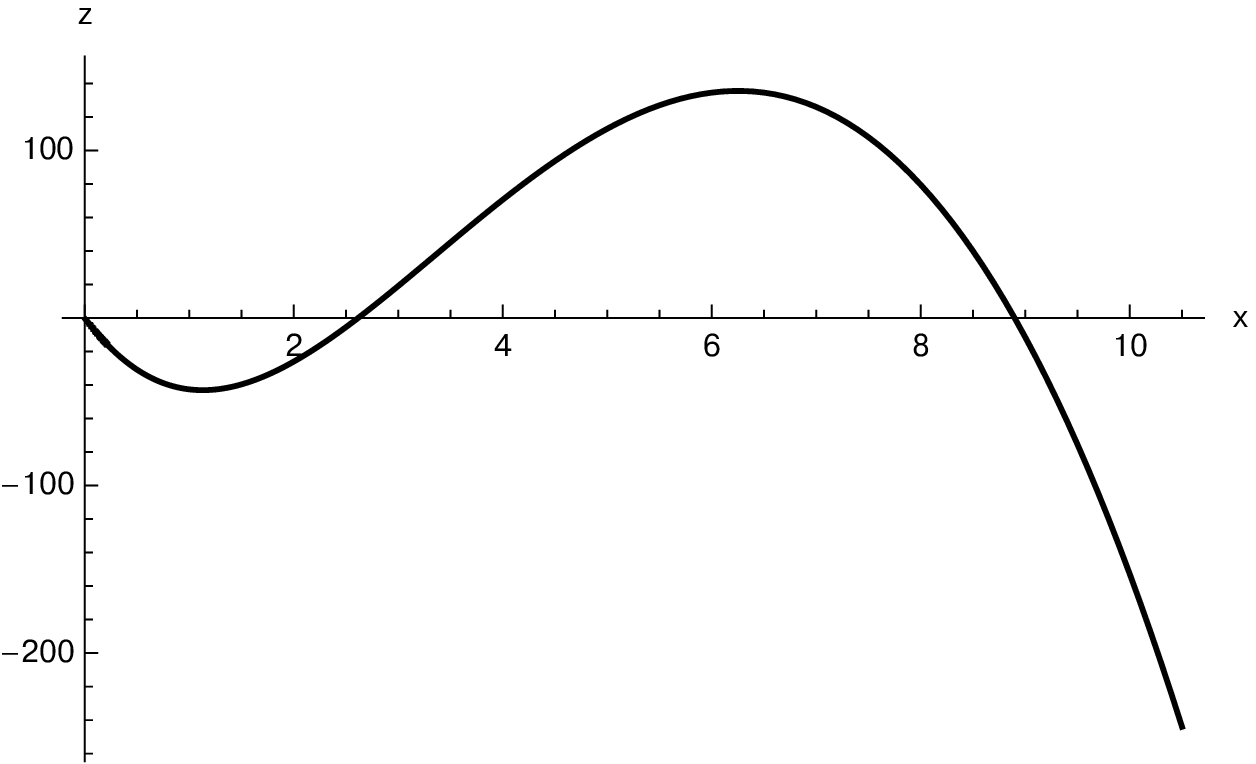}
\end{center}
  \caption{
Left panel: comparison of $\color{blue} \dd\calr=\calr^{\caln=2^*}_{(\mu,i\mu H/2)}-\calr_{loc}^0$ (blue curve) and
the best two parameter
fit $\color{red} \dd\calr=\calr_{loc}^{ambiguity}$ (red curve).
Right panel: the residual $\color{black} \dd\calr-\calr_{loc}^{ambiguity}$ (black curve).
} \label{figure8}
\end{figure}

In fig.~\ref{figure8} we attempted to adjust the coefficients $\{\a_1,\a_2\}$ to match
$\calr_{loc}$ and $\calr^{\caln=2^*}_{(\mu,i\mu H/2)}$ computed in section \ref{mainmuimu}.
The blue curve (left panel) represents
\begin{equation}
\dd\calr=\calr^{\caln=2^*}_{(\mu,i\mu H/2)}-\calr_{loc}^{0}\,,
\eqlabel{defddr}
\end{equation}
and the right curve (left panel) represents the best fit to \eqref{defddr} using \eqref{rscheme}
with $\a_0=0$. The residual of the best fit is shown in the right panel. Clearly,
the interpretation for
\begin{equation}
s_{ent}=s_{ent,loc}
\eqlabel{wrong2}\,,
\end{equation}
attempted in this section, is incorrect.

\section*{Acknowledgments}
Research at Perimeter
Institute is supported by the Government of Canada through Industry
Canada and by the Province of Ontario through the Ministry of
Research \& Innovation. This work was further supported by
NSERC through the Discovery Grants program.

\appendix
\section{BEFP equations of motion}
Within the ansatz \eqref{efbmetric}, we obtain the following evolution  and the
constraint equations from \eqref{sbefp}:
\begin{equation}
\begin{split}
&0=\left(d_+\Sigma\right)'+2 {\Sigma'}\ d_+\ln\Sigma-\frac{1}{\Sigma}+
\frac \Sigma6\ V_{BEFP},\\ 
&0=A''-6(\ln\Sigma)'\ d_+\ln\Sigma +\frac{3}{\Sigma^2}+\frac{2(\bar{z}' d_+ z+{z}' d_+\bar{z})}{(1-z\bar{z})^2} +12\eta' d_+\eta 
-\frac {V_{BEFP}}{6}, \\ 
&(d_+\eta)'+\left(\frac{3}{2}(\ln\Sigma)'-(\ln\eta)'\right)d_+\eta+\frac 32\eta' d_+\ln\Sigma  
-\frac {\eta^2}{48} \del_\eta V_{BEFP}, \\ 
&0=(d_+\bar{z})'+\left(\frac{2z\bar{z}'}{1-z\bar{z}}+\frac{3}{2} (\ln\Sigma)'\right) d_+\bar{z}
+\frac 32\bar{z}' d_+\ln\Sigma -\frac{(1-z\bar{z})^2}{8}\del_z V_{BEFP},
\\ 
&0=(d_+{z})'+\left(\frac{2\bar{z}{z}'}{1-z\bar{z}}+\frac{3}{2} (\ln\Sigma)'\right) d_+{z}
+\frac 32{z}' d_+\ln\Sigma -\frac{(1-z\bar{z})^2}{8}\del_{\bar{z}} V_{BEFP},
\end{split}
\eqlabel{evb}
\end{equation}
\begin{equation}
0=\Sigma''+4\left( \frac{(\eta')^2}{\eta^2}+\frac {z'\bar{z}'}{3(1-z\bar{z})^2}\right) \Sigma\,,
\eqlabel{hamb}
\end{equation}
\begin{equation}
\begin{split}
&0=d^2_+\Sigma-A'd_+\Sigma+4 \left(\frac{(d_+\eta)^2}{\eta^2}
+\frac {d_+ zd_+\bar{z}}{3(1-z\bar{z})^2}\right)\Sigma  \,.
\end{split}
\eqlabel{momb}
\end{equation}
Taking the late-time limit, and using \eqref{vacn2*}, we find:
\begin{equation}
\begin{split}
&0=\eta_v''-\frac{(\eta_v')^2}{\eta_v}
+\left(\frac{3H}{2A_v}+\left(\ln A_v\sigma_v^3\right)'\right)\eta_v'-\frac{\eta_v^2}{48 A_v}\del_\eta V_{BEFP}\,,\\
&0=z_v''+\frac{2\bar{z}_v(z_v')^2}{1-z_v\bar{z}_v}+\left(\frac{3H}{2A_v}+\left(\ln A_v\sigma_v^3\right)'\right)z_v'-\frac{(1-z_v\bar{z}_v)^2}{8 A_v}\del_{\bar{z}} V_{BEFP}\,,\\
&0=\bar{z}_v''+\frac{2{z}_v(\bar{z}_v')^2}{1-z_v\bar{z}_v}+\left(\frac{3H}{2A_v}+\left(\ln A_v\sigma_v^3\right)'\right)\bar{z}_v'-\frac{(1-z_v\bar{z}_v)^2}{8 A_v}\del_{{z}} V_{BEFP}\,,\\
&0=\sigma_v''+\frac 43 \sigma_v\left(3\frac{(\eta_v')^2}{\eta_v^2}+\frac{z_v'\bar{z}_v'}
{(1-z_v\bar{z}_v)^2}\right)\,,\\
&0=A_v''+4 A_v\left(3\frac{(\eta_v')^2}{\eta_v^2}+\frac{z_v'\bar{z}_v'}{(1-z_v\bar{z}_v)^2}\right)
-6 A_v \left((\ln \sigma_v)'\right)^2-6 H(\ln\sigma_v)'
-\frac {1}{6}V_{BEFP}\,,
\end{split}
\eqlabel{veoms}
\end{equation}
along with the constraints
\begin{equation}
\begin{split}
&0=\sigma_v'+\frac{\sigma_v}{2A_v}(H-A_v')\,,\\
&0=\frac{(\eta_v')^2}{\eta_v^2}+\frac{z_v'\bar{z}_v'}{3(1-z_v\bar{z}_v)^2}
-\frac 12 \left((\ln\sigma_v)'\right)^2-\frac{\sigma_v'}{4\sigma_vA_v}(3H+A_v')
-\frac{1}{24A_v} V_{BEFP}\,.
\end{split}
\eqlabel{veoms2}
\end{equation}
It is straightforward to verify that constraints \eqref{veoms2} are consistent with \eqref{veoms}.

\section{$(m,k)\ =\ \left(i\mu\,,\ \mu/2\right)$ equations of motion}\label{case1disc}

In FG coordinate system \eqref{deffg1} the gravitational equations of motion take form:
\begin{equation}
\begin{split}
0=&z_v'+\frac{\sqrt{(\eta_v^6 \bar{z}_v^2-\eta_v^6+\bar{z}_v^2+1) (z_v^2 \eta_v^6-\eta_v^6+z_v^2+1)}(\eta_v^6 z_v-\eta_v^6 \bar{z}_v+2 z_v+2 \bar{z}_v)}{2\eta_v^2 x (\eta_v^6 \bar{z}_v^2-\eta_v^6+\bar{z}_v^2+1)}\,,\\
0=&\bar{z}_v'-\frac{\sqrt{(\eta_v^6 \bar{z}_v^2-\eta_v^6+\bar{z}_v^2+1)
(z_v^2 \eta_v^6-\eta_v^6+z_v^2+1)} (\eta_v^6 z_v-\eta_v^6 \bar{z}_v-2 z_v-2 \bar{z}_v)}{2\eta_v^2 x (z_v^2 \eta_v^6-\eta_v^6+z_v^2+1)}\,,\\
0=&\eta_v'+ \frac{\sqrt{(\eta_v^6 \bar{z}_v^2-\eta_v^6+\bar{z}_v^2+1)
(z_v^2 \eta_v^6-\eta_v^6+z_v^2+1)}}{3\eta_v (z_v \bar{z}_v-1) x}\,,\\
0=&s_v'-\frac{(\eta_v^6 \bar{z}_v^2-\eta_v^6-2 \bar{z}_v^2-2) (1+x s)
\sqrt{(\eta_v^6 \bar{z}_v^2-\eta_v^6+\bar{z}_v^2+1) (\eta_v^6 z_v^2-\eta_v^6+z_v^2+1)}}
{3x^2 \eta_v^2 (z_v \bar{z}_v-1) (\eta_v^6 \bar{z}_v^2-\eta_v^6+\bar{z}_v^2+1)}-\frac{1}{x^2}\,.
\end{split}
\eqlabel{case1r1}
\end{equation}
This coordinate system covers the first holographic bulk
subregion (see discussion in section \ref{split}) --- from the asymptotic $AdS$ boundary to $A_v=0$.

\section{$(m,k)\ =\ \left(\mu\,,\ \mu/2\right)$ equations of motion}\label{physmass}
In FG coordinate system \eqref{defg3} the gravitational equations of motion take form:
\begin{equation}
\begin{split}
0=&z_{1,v}''-\frac{2 z_{1,v} ((z_{1,v}')^2-(z_{2,v}')^2)}{z_{2,v}^2+z_{1,v}^2-1}
-\frac{4 z_{2,v} z_{1,v}' z_{2,v}'}{z_{2,v}^2+z_{1,v}^2-1}
+\frac{2z_{1,v}'}{x}\\
&+\frac{h^{1/2} \eta_v^2 z_{1,v} (z_{2,v}^2 (\eta_v^6+2)+2 z_{1,v}^2-2)}{2x^2 (z_{2,v}^2+z_{1,v}^2-1)}
+\frac{5 z_{1,v}' \Theta^{1/2}}{6h x \eta_v^2 (z_{2,v}^2+z_{1,v}^2-1)}\,,\\
0=&z_{2,v}''+\frac{2 z_{2,v} ((z_{1,v}')^2-((z_{2,v}')^2)}{z_{2,v}^2+z_{1,v}^2-1}
-\frac{4 z_{1,v} z_{1,v}' z_{2,v}'}{z_{2,v}^2+z_{1,v}^2-1}+\frac{2 z_{2,v}'}{x}\\
&+\frac{\eta_v^2 z_{2,v} h^{1/2} (\eta_v^6 (z_{2,v}^2-z_{1,v}^2+1)+4 z_{2,v}^2+4 z_{1,v}^2-4)}
{4 x^2 (z_{2,v}^2+z_{1,v}^2-1)}+\frac{5z_{2,v}' \Theta^{1/2}}{6h x \eta_v^2 (z_{2,v}^2+z_{1,v}^2-1)}\,,\\
0=&\eta_v''-\frac{(\eta_v')^2}{\eta_v}+\frac{2 \eta_v'}{x}
+\frac{5\eta_v' \Theta^{1/2}}{6h \eta_v^2 (z_{2,v}^2+z_{1,v}^2-1) x}-\frac{h^{1/2}
}{6x^2 \eta_v^3 (z_{2,v}^2+z_{1,v}^2-1)^2}\biggl(2 z_{2,v}^2 \eta_v^{12}\\&+((z_{2,v}^2+z_{1,v}^2)^2-1)
\eta_v^6+(z_{2,v}^2+z_{1,v}^2-1)^2\biggr)\,,\\
0=&h'+\frac{4 h}{x}+\frac{2\Theta^{1/2}}{3\eta_v^2 (z_{2,v}^2+z_{1,v}^2-1) x}\,,\\
0=&s'-h^{1/2} s\,,
\end{split}
\eqlabel{case3r1}
\end{equation}
where
\begin{equation}
\begin{split}
\Theta=&12 h^2 x^2 \eta_v^2 \biggl(3 (z_{2,v}^2+z_{1,v}^2-1)^2 (\eta_v')^2
+\eta_v^2 ((z_{1,v}')^2+(z_{2,v}')^2)\biggr)\\
&+36 \eta_v^4 x^2
(z_{2,v}^2+z_{1,v}^2-1)^2 h^3-3 h^{5/2} \biggl(z_{2,v}^2 \eta_v^{12}+2 ((z_{2,v}^2+z_{1,v}^2)^2-1) \eta_v^6
\\&-(z_{2,v}^2+z_{1,v}^2-1)^2\biggr)\,.
\end{split}
\eqlabel{deftheta}
\end{equation}
This coordinate system covers the first holographic bulk
subregion (see discussion in section \ref{split}) --- from the asymptotic $AdS$ boundary to $h=0$.

\section{IR thermodynamics of $\caln=2^*$ plasma}\label{IRthermodynamics}

The ratio of the bulk viscosity to the shear viscosity in $\caln=2^*$ plasma
saturates the bulk viscosity bound \cite{Buchel:2007mf} in the deep IR, \ie
\begin{equation}
\lim_{T/m\to 0}\ \left[\frac{\zeta}{\eta}-2\left(\frac 13-c_s^2\right)\right]\approx 0\,,
\eqlabel{bvbound}
\end{equation}
as
\begin{equation}
\lim_{T/m\to 0}\ \left(\frac 13-c_s^2\right) \approx 0.08\approx \frac{1}{12}\qquad \Longrightarrow\qquad  c_s^2\to \frac{1}{4}\,.
\eqlabel{sat}
\end{equation}
Since the bulk viscosity bound is automatically saturated for Kaluza-Klein (KK) reductions of
higher dimensional conformal gauge theories to three spatial dimensions \cite{Buchel:2007mf}, \eqref{bvbound} and \eqref{sat}
strongly suggests that $\caln=2^*$ IR thermodynamics is a KK reduction of that of emergent $CFT_5$\footnote{Recall that in
$CFT_d$, $c_s^2=\frac{1}{d-1}$.}. This is indeed the case \cite{HoyosBadajoz:2010td}.

Following \cite{HoyosBadajoz:2010td}, consider the $\caln=2^*$ vacuum in a holographic dual, the PW geometry \cite{Pilch:2000ue}.
The IR limit corresponds to $\chi\to \infty$, thus, introducing a new radial coordinate $u\to\infty$,
\begin{equation}
e^{2\chi}\simeq 2u\,,\qquad e^{6\a}\simeq \frac{2}{3u}\,,\qquad e^A\simeq\left(\frac{2}{3u^4}\right)^{1/3} {\rm k}\,.
\eqlabel{change}
\end{equation}
the background metric becomes (we set the five-dimensional supergravity coupling $g=1$)
\begin{equation}
ds_{PW}^2\simeq\left(\frac{3}{2u^2}\right)^{4/3}\left[4 du^2 +\left(\frac{2{\rm k}}{3}\right)^2 \eta_{\mu\nu}dx^\mu dx^\nu\right]\,.
\eqlabel{irmetric}
\end{equation}
The parameter ${\rm k}$ here is defined as in PW \cite{Pilch:2000ue}.
Introducing \cite{HoyosBadajoz:2010td}
\begin{equation}
e^{4\phi_2}\equiv e^{2(\a-\chi)}\simeq \left(\frac{1}{12u^4}\right)^{1/3}\,,\qquad e^{4\phi_1}\equiv e^{6\a+2\chi}\simeq \frac 43\,,
\eqlabel{phi12}
\end{equation}
the metric \eqref{irmetric} can be understood as a KK reduction of the locally $AdS_6$ metric
on a compact $x_6\sim x_6 +L_6$:
\begin{equation}
ds_6^2=e^{-2\phi_2} ds_{PW}^2+e^{6\phi_2} dx_6^2\,\,\simeq\,\, \frac{3^{3/2}}{2u^2}
\left[4 du^2+\left(\frac{2{\rm k}}{3}\right)\eta_{\mu\nu}dx^\mu dx^\nu+\frac 19 dx_6^2\right]\,.
\eqlabel{6metric1}
\end{equation}
The metric \eqref{6metric1} and the scalar $\phi_1$ \eqref{phi12} is a solution \cite{HoyosBadajoz:2010td}
to $d=6$ $\caln=(1,1)$  $F(4)$ SUGRA \cite{Romans:1985tw}
\begin{equation}
S_{F(4)}=\frac{1}{16\pi G_6}\int_{\calm_6} d\xi^6 \sqrt{-g_6} \left(R_6-4(\del\phi_1)^2+e^{-2\phi_1}+e^{2\phi_1}-\frac 16e^{6\phi_1}\right)\,,
\eqlabel{action6}
\end{equation}
where, using the PW five-dimensional Newton's constant $G_5$,
\begin{equation}
\frac{L_6}{G_6}=\frac{1}{G_5}=\frac{N^2}{4\pi}\,.
\eqlabel{defg6}
\end{equation}

The IR thermodynamics of $\caln=2^*$ plasma is thus the (appropriately rescaled) thermodynamics of AdS-Schwarzschild black branes
in \eqref{action6}. Specifically,
\begin{equation}
\left(ds_6^{BH}\right)^2=\frac{3^{3/2}}{2u^2}\left(-\left(1-\frac{u^5}{u_0^5}\right)(d{\hat t})^2+d\hat{\boldsymbol{x}}^2
+(d\hat{x}_6)^2
+4\left(1-\frac{u^5}{u_0^5}\right)^{-1} (du)^2\right)\,,
\eqlabel{bh6}
\end{equation}
where the rescaled, \ie $\hat{\ }$ coordinates, are related to PW coordinates $x^\mu$ and the KK direction $x_6$
as follows (compare with \eqref{6metric1}):
\begin{equation}
\{{\hat t},{\hat{\boldsymbol{x}}}\}\equiv {\hat x}^\mu=\frac {2{\rm k}}{3} x^\mu\,,\qquad {\hat x}_6=\frac 13 x_6\,.
\eqlabel{rescale}
\end{equation}
The black brane \eqref{bh6} Hawking temperature (conjugate to the PW time coordinate) is
\begin{equation}
T=\frac{5{\rm k}}{12\pi u_0}\,. 
\eqlabel{tu0}
\end{equation}
The entropy density (per unit PW volume) of the black brane \eqref{bh6} is
\begin{equation}
s=\frac{1}{4 G_6}\ \frac{27}{4u_0^4}\ L_6\ \left(\frac{2{\rm k}}{3}\right)^3\ \frac 13
=\frac{1}{4G_5}\ \frac{13824\pi^4 T^4}{625 {\rm k}}=\frac{6912\pi^4}{625}\
\left(\frac{m^2}{T^2}\right)^{-1/2}\ \frac{T^3}{4 G_5}\,,
\end{equation}
where we used the identification \cite{Buchel:2000cn} ${\rm k}=2m$.
Thus, the deep IR entropy density of $\caln=2^*$ plasma scales as 
\begin{equation}
\frac{4 G_5 s}{T^3}\ \to\ \frac{6912\pi^4}{625}\
\left(\frac{m^2}{T^2}\right)^{-1/2}\qquad {\rm as}\qquad \frac{m^2}{T^2}\to \infty\,.
\eqlabel{scaling}
\end{equation}

\bibliographystyle{JHEP}
\bibliography{n2desitter}

\end{document}